\documentclass[pdflatex,sn-mathphys-num]{sn-jnl}

\usepackage{graphicx}%
\usepackage{multirow}%
\usepackage{amsmath,amssymb,amsfonts}%
\usepackage{amsthm}%
\usepackage{mathrsfs}%
\usepackage[title]{appendix}%
\usepackage{xcolor}%
\usepackage{textcomp}%
\usepackage{manyfoot}%
\usepackage{booktabs}%
\usepackage{algorithm}%
\usepackage{algorithmicx}%
\usepackage{algpseudocode}%
\usepackage{listings}%

\usepackage{placeins} 
\usepackage{caption}
\usepackage{chngcntr}  
\usepackage{subcaption}
\usepackage{array}      
\usepackage{lscape}     
\usepackage{graphicx}   
\usepackage{ragged2e}   
\usepackage{makecell}
\usepackage{bm}
\usepackage{lineno} 
\usepackage{color}
\usepackage{pgffor}
\usepackage[T1]{fontenc}

\theoremstyle{thmstyleone}%
\theoremstyle{thmstyletwo}%

\theoremstyle{thmstylethree}%

\begin{document}

\title[``Recognition,'' ``belief,'' and ``action'' regarding conspiracy theories]{The ``recognition,'’ ``belief,'' and ``action'' regarding conspiracy theories: An empirical study using large-scale samples from Japan and the United States}


\author[1]{\fnm{Taichi} \sur{Murayama}}\email{murayama-taichi-bs@ynu.ac.jp}

\author[2]{\fnm{Dongwoo} \sur{Lim}}\email{dongwoo.lim@tsuda.ac.jp}

\author[3]{\fnm{Akira} \sur{Matsui}}\email{matsui-akira-zr@ynu.ac.jp}

\author*[4]{\fnm{Tsukasa} \sur{Tanihara}}\email{tanihara@fc.ritsumei.ac.jp}

\affil*[1]{\orgdiv{Faculty of Environment and Information Sciences, Division of Social Environment and Information}, \orgname{Yokohama National University}, \orgaddress{\state{Kanagawa}, \country{Japan}}}

\affil[2]{\orgdiv{Department of Policy Studies}, \orgname{Tsuda University}, \orgaddress{\state{Tokyo}, \country{Japan}}}

\affil[3]{\orgdiv{Division of International Social Sciences, Faculty of International Social Sciences}, \orgname{Yokohama National University}, \orgaddress{\state{Kanagawa}, \country{Japan}}}

\affil[4]{\orgdiv{Department of Social Science}, \orgname{Ritsumeikan University}, \orgaddress{\state{Kyoto}, \country{Japan}}}

\abstract{
Conspiracy theories present significant societal challenges, shaping political behavior, eroding public trust, and disrupting social cohesion. 
Addressing their impact requires recognizing that conspiracy engagement is not a singular act but a multi-stage process involving distinct cognitive and behavioral transitions. 
In this study, we investigate this sequential progression, ``recognition,'' ``belief,'' and ``action'' (demonstrative action and diffusion action), using nationally representative surveys from the United States (N=13,578) and Japan (N=16,693). 
Applying a Bayesian hierarchical model, we identify the key social, political, and economic factors that drive engagement at each stage, providing a structured framework for understanding the mechanisms underlying conspiracy theory adoption and dissemination.
We find that recognition serves as a crucial gateway determining who transitions to belief, and that demonstrative and diffusion actions are shaped by distinct factors.
Demonstrative actions are more prevalent among younger, higher-status individuals with strong political alignments, whereas diffusion actions occur across broader demographics, particularly among those engaged with diverse media channels.
Our findings further reveal that early-life economic and cultural capital significantly influence the shape of conspiratorial engagement, emphasizing the role of life-course experiences.
These insights highlight the necessity of distinguishing between different forms of conspiracy engagement and highlight the importance of targeted interventions that account for structural, cultural, and psychological factors to mitigate their spread and societal impact.
}




\maketitle


\section*{Introduction}\label{intro}

Conspiracy theories have emerged as an increasingly significant issue in understanding social and political events. 
Conspiracy theories are defined as ``attempts to explain the ultimate causes of significant social and political events and circumstances with claims of secret plots by two or more powerful actors~\cite{douglas2019understanding}'' or ``an explanation for important events and circumstances that involve secret plots by groups with malevolent agendas~\cite{douglas2017psychology}.''
Historically, prominent conspiracy theories have been associated with events such as the 9/11 terrorist attacks~\cite{swami2010unanswered}, the death of Princess Diana~\cite{douglas2008hidden}, and the assassination of John F. Kennedy~\cite{mccauley1979popularity}.
More recently, the global spread of COVID-19 has triggered a surge in conspiracy theories, such as claims that ``COVID-19 is part of a government bioweapons program, that 5G cell towers are spreading COVID-19, and that pharmaceutical companies are encouraging the spread of COVID-19 for profit~\cite{earnshaw2020covid,douglas2021covid,van2022antecedents}.''
These conspiracy beliefs are part of an effort by individuals to make sense of a chaotic world, with large-scale societal events often acting as triggers for the emergence of new conspiracy theories~\cite{franks2017beyond}.

In particular, the rapid development of the internet and social media has unprecedentedly accelerated the spread of conspiracy beliefs. 
While the online environment offers the convenience of disseminating information quickly and broadly, it also provides fertile ground for conspiracy theories to proliferate and spread~\cite{sunstein2009conspiracy,miller2016conspiracy}. 
Phenomena such as echo chambers and filter bubbles skew users' perceptions, while deliberate disruptions by trolls and the use of automated accounts, such as bots, further facilitate the dissemination of conspiracy theories. 
Consequently, individuals are more likely to accept misinformation, disinformation, or misleading information as facts, amplifying their influence~\cite{finkelstein2020covid,hoseini2023globalization,islam2020covid,marchal2020covid19, rovetta2020global}.

The proliferation of conspiracy beliefs has serious adverse effects on society. 
Specifically, it fosters prejudice and discrimination~\cite{jolley2020exposure,tangherlini2020automated}, undermines democratic norms~\cite{albertson2020conspiracy}, increases political apathy~\cite{jolley2014social}, reduces trust in governments and public institutions~\cite{lutkenhaus2019mapping,imhoff2020bioweapon}, leads to non-compliance with public health guidelines~\cite{hornsey2018psychological,pummerer2022conspiracy,jolley2014effects}, and even escalates involvement in crime and riots~\cite{jolley2019belief}. 
These impacts deepen societal divisions, hinder collective action and consensus-building, and pose significant risks to social stability~\cite{islam2020covid,lowicki2022does}.

Numerous studies have examined the characteristics of individuals who hold conspiracy beliefs. 
For instance, Douglas et al.~\cite{douglas2017psychology} argue belief in conspiracy theories appears to be driven by motives that can be characterized as epistemic (understanding one’s environment), existential (being safe and in control of one’s environment), and social (maintaining a positive image of the self and the social group.)
Beyond psychological traits~\cite{hornsey2023individual, stasielowicz2022believes, bowes2023conspiratorial, biddlestone2022systematic}, attributes such as age and gender~\cite{enders2024sociodemographic, hettich2022conspiracy}, social factors like educational attainment and political ideology~\cite{imhoff2014speaking, alper2021psychological,sallam2020covid}, and connections to specific cultural backgrounds or group affiliations~\cite{van2021cultural} have all been suggested to correlate with conspiracy beliefs.

Previous studies have primarily focused on the ``belief'' in conspiracy theories. 
However, as a prerequisite for belief, individuals first ``recognize'' conspiracy theories. 
Additionally, attention must also be paid to demonstrative actions stemming from conspiracy theories; the most extreme example in recent history is the 2021 Capitol attack in the United States~\cite{armaly2022christian}.
Nevertheless, only a small fraction of those who hold such beliefs escalate to participating in social movements or acts of violence~\cite{jolley2019belief,imhoff2014speaking}. 
These observations indicate that the extent and forms of influence conspiracy theories exert on individuals vary, necessitating an examination of how social attributes shape these differences.

This study focuses on the three stages of conspiracy theories, ``recognition,'' ``belief,'' and ``action,'' in the contexts of Japan and the United States. 
It aims to explore the relationships between each stage and individuals’ attributes as well as their social strata. 
By examining whether these relationships align with the characteristics of individuals holding conspiracy beliefs as indicated in prior research or whether new factors and mechanisms emerge, this study seeks to provide a comprehensive understanding of the spread of conspiracy theories and their social impacts.


\section*{Theoretical Background and Variable Settings}\label{background}
Numerous factors influencing the formation of conspiracy beliefs have been examined in previous research. 
Building on these findings, this paper explores how these factors operate across three key stages: the initial ``recognition,'' their subsequent ``belief,'' and their eventual translation into ``action.''
This section introduces the main elements under consideration.

\subsection*{Individual Attributes}
Previous research on how individual attributes, such as gender and age, affect conspiracy beliefs has yielded inconsistent findings.
In terms of gender, studies conducted in Turkey, Jordan, Kuwait, Saudi Arabia, and Greece suggest that women are more likely to believe in conspiracy theories compared to men~\cite{van2020engagement, alper2021psychological, sallam2020covid, sallam2021high}.
In contrast, Cassese et al. report that men in the United States exhibit a tendency to believe such theories~\cite{cassese2020gender}, while other investigations reveal no significant gender differences in believing COVID-19 conspiracy claims~\cite{earnshaw2020covid, freeman2022coronavirus, kuhn2022coronavirus}.
Age-related findings are similarly inconsistent. 
Numerous studies show that younger individuals are more likely to believe in conspiracy theories~\cite{freeman2022coronavirus,kuhn2022coronavirus,uscinski2020people,enders2024sociodemographic,goreis2019systematic, de2021beliefs, romer2020conspiracy, juanchich2021covid}, potentially due to higher social media usage, lower self-esteem, and dissatisfaction with politics~\cite{bordeleau2024relationship}. 
Conversely, studies involving samples from Brazilian and Portuguese suggests that older individuals are more likely to believe COVID-19-related conspiracy theories~\cite{van2020engagement}.
Van Mulukom et al. highlight inconsistency in the the relationship between age and conspiracy beliefs, suggesting that cultural and social factors may play a significant role in shaping the relationship~\cite{van2022antecedents}.
These conflicting results imply that the relationship between individual attributes and conspiracy beliefs may vary depending on cultural and social contexts.

\subsection*{Social Status}
The impact of income and economic status on the formation of conspiracy beliefs has been widely discussed in research. 
In this context, attention should also be given to the role of employment type, particularly the influence of stable permanent employment.
Studies suggest that lower income is a contributing factor to stronger conspiracy beliefs~\cite{enders2024sociodemographic, hettich2022conspiracy, constantinou2021covid, duplaga2020determinants}. 
Individuals in precarious employment or those who perceive themselves as having low social status are often more likely to accept conspiracy theories, as a way to rationalize their dissatisfaction and perceived societal inequalities~\cite{casara2022impact, jetten2022economic}. 
Plenta argues that individuals with lower socioeconomic status may adopt conspiracy theories to attribute their hardships to more successful individuals~\cite{plenta2020conspiracy}.
This perspective suggests that subjective perceptions of economic inequality are one of the mechanisms that reinforce conspiracy beliefs. 
Conversely, stable permanent employment may provide a sense of economic stability and social status, potentially contributing to the suppression of conspiracy beliefs.

\subsection*{Educational Attainment}
Many studies have reported that lower levels of education are associated with a higher tendency to believe in conspiracy theories~\cite{enders2024sociodemographic, hettich2022conspiracy, constantinou2021covid, duplaga2020determinants, casara2022impact, jetten2022economic, plenta2020conspiracy, douglas2016someone, van2017education, swami2011recognition}.
Lower educational attainment has been linked to a lack of critical thinking and analytical reasoning skills, which may contribute to an increase in conspiracy beliefs~\cite{bago2022does, swami2014analytic}.
However, Galliford \& Furnham (2017) found no significant relationship between educational attainment and beliefs in political conspiracy theories~\cite{galliford2017individual}.
Additionally, Roscigno (2024) suggested that the relationship between educational attainment and conspiracy beliefs may follow a curvilinear distribution~\cite{roscigno2024status}. 
This implies that even highly educated individuals may exhibit a tendency to believe in conspiracy theories, particularly when influenced by factors such as social media use. 
These findings indicate that the impact of educational attainment on conspiracy beliefs is not straightforward and involves complex interactions with the environment.

\subsection*{Political Attitudes}
The relationship between political attitudes and conspiracy beliefs has been a key focus in numerous studies. 
Some research suggests that right-wing political attitudes are strongly associated with conspiracy beliefs~\cite{miller2016conspiracy, djordjevic2021beyond, grzesiak2009right, van2021paranoid}. 
This may be influenced by the preference among conservatives for echo-chamber environments that reinforce existing worldviews~\cite{jost2018ideological}.
However, many studies have reported that the relationship follows a curvilinear pattern, where both extreme right-wing and left-wing positions are more strongly linked to conspiracy beliefs compared to moderate political stances~\cite{sutton2020conspiracy,imhoff2022conspiracy,van2015political,nera2021power,krouwel2017does}.
This U-shaped pattern may be attributed to the tendency of individuals with extreme political ideologies to reject opposing groups and narratives, which fosters a susceptibility to conspiracy beliefs~\cite{brandt2014ideological,cichocka2016grammar,van2017extreme}. 
These studies suggest that political extremism is associated with the development and reinforcement of conspiracy beliefs in many countries.

\subsection*{Trust in Government}
Existing research consistently shows that a decline in trust in governments or medical institutions strengthens the tendency to rely on conspiratorial information~\cite{lutkenhaus2019mapping,imhoff2020bioweapon}. 
Additionally, a lack of political control and dissatisfaction with the political system have been suggested as factors that amplify conspiracy beliefs~\cite{jolley2020pylons,edelson2017effect,nyhan2017media,kofta2020breeds}.
When the political environment fails to meet individuals’ expectations, heightened anxiety and uncertainty may arise, motivating individuals to adopt conspiratorial interpretations as a way to make sense of their situation.

\subsection*{Trust in Science}
Distrust in science is a major driver of conspiracy beliefs, largely because it contributes to the spread of misinformation and the rejection of expert consensus. 
This effect is especially pronounced in public health-related conspiracy theories, such as those surrounding COVID-19 vaccines and infection control measures~\cite{vranic2022did, van2020engagement, brzezinski2020belief, roozenbeek2020susceptibility}.
A decline in trust in science is also linked to denialism, which involves rejecting scientific discourse based on overconfidence and misperceived self-knowledge. Denialism is considered a key predictor of conspiracy beliefs~\cite{uscinski2020people}.

\subsection*{Religiosity}
Conspiracy theories often exhibit religious characteristics, and it has been suggested that the two may be connected through common psychological mechanisms~\cite{franks2013conspiracy}. 
Religious individuals are more likely to perceive world events as being orchestrated by an intentional force rather than as random or coincidental, making them more susceptible to conspiracy beliefs~\cite{imhoff2014speaking,van2017education}. 
This cognitive tendency serves as a common foundation for the supernatural worldview of religious beliefs and the interpretive framework of conspiracy theories.
Empirical studies have consistently demonstrated a positive correlation between religiosity and conspiracy beliefs~\cite{galliford2017individual, lowicki2022does, lobato2014examining,frenken2023relation}. 
Jasinskaja \& Jetten (2019) found that not only does religiosity strengthen conspiracy beliefs, but this relationship is partially mediated by anti-intellectualism~\cite{jasinskaja2019unpacking}. 
These findings suggest that religious worldviews may encourage a rejection of scientific and rational interpretations, thereby fostering susceptibility to conspiracy beliefs.

\subsection*{Media Usage}
The type of media individuals consume plays a crucial role in shaping their susceptibility to conspiracy beliefs. 
In general, greater exposure to traditional media is associated with lower levels of belief in conspiracy theories and misinformation, whereas exposure to digital media tends to reinforce such beliefs~\cite{freeman2022coronavirus, kuhn2022coronavirus, de2021beliefs, romer2020conspiracy, chen2021effects}. 
For example, traditional media exposure has been linked to reduced belief in conspiracy theories in countries such as Belgium, Switzerland, and Canada~\cite{de2021beliefs}. 
Conversely, Boberg et al. (2020) highlighted the role of social media in increasing exposure to alternative news sources, which play a significant role in amplifying the spread of conspiracy theories~\cite{boberg2020pandemic}.
These findings underscore the importance of media consumption patterns in shaping public perceptions of conspiracy theories.

\subsection*{Social Capital}
Although research on the influence of social capital on conspiracy beliefs is limited, it is considered a significant variable. 
When examining the profiles of individuals involved in conspiracy theories, the social stratification stemming from their upbringing offers crucial insights into the societal positioning of conspiracy theories.
For example, Callaghan et al. (2019) found that parental conspiracy thinking is associated with delays in vaccinating their children, highlighting the intergenerational transmission of conspiratorial thinking~\cite{callaghan2019parent}. 
Similarly, Freeman \& Bentall (2017) reported that individuals who support conspiracy beliefs are more likely to have experienced disruptive family environments during childhood, such as not living with their parents or being exposed to violence~\cite{freeman2017concomitants}. 
These studies suggest that childhood family environments and the resultant social capital may influence the formation of conspiracy beliefs. 
Casara et al. (2022), using the Gini coefficient as a measure of inequality, demonstrated that both objective economic disparity and subjective perceptions of inequality are correlated with conspiracy beliefs~\cite{casara2022impact}. 
Their experiment revealed that individuals who perceive themselves as being in economic inequality are more likely to support conspiratorial narratives. 
These results underscore the complex interplay between social capital, economic conditions, and the development of conspiracy beliefs.

\section*{Results}

We examine how individuals engage with conspiracy theories across three distinct stages, recognition, belief, and action, and investigate how each stage is influenced by a variety of personal and social factors, including demographic characteristics, social status, political orientation, media usage habits, and social capital. 
This approach reflects the premise that conspiracy theory engagement is not a singular act but rather a multi-step process. 
Individuals must first recognize a conspiracy theory, believe it once recognized, and only then may they act upon that belief by demonstrating in public or actively disseminating the theory on social media.
To capture both the direct effects of these explanatory variables at each stage and the indirect effects that transmit through prior stages, we employ a Bayesian Hierarchical Bernoulli model. 
Further details of this modeling framework and the description of explanatory variables are provided in \nameref{method}.

\subsection*{Survey Overview}
We initially collected 19,783 responses in the U.S. and 20,150 in Japan. After quality control and data cleaning, the final analyzed samples consisted of 13,578 respondents from the U.S. and 16,693 from Japan. 
Respondents provided demographic information, such as social status, political orientation, media usage habits, and social capital, as well as their recognition, belief, and actions regarding a set of conspiracy theories.

\begin{figure}[t]
    \centering
    \begin{subfigure}{0.45\linewidth}
        \centering
        \includegraphics[width=\linewidth]{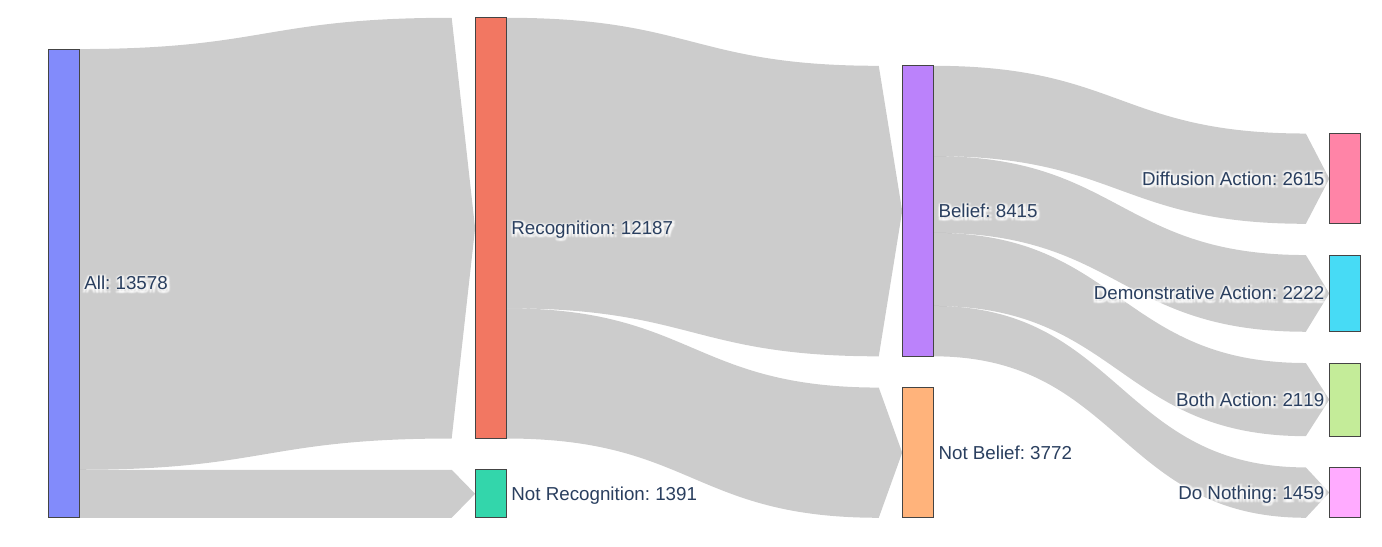}
        \caption{Sankey diagram showing the sequential stages of conspiracy theory engagement in the U.S.}
    \end{subfigure}
    \hfill
    \begin{subfigure}{0.45\linewidth}
        \centering
        \includegraphics[width=\linewidth]{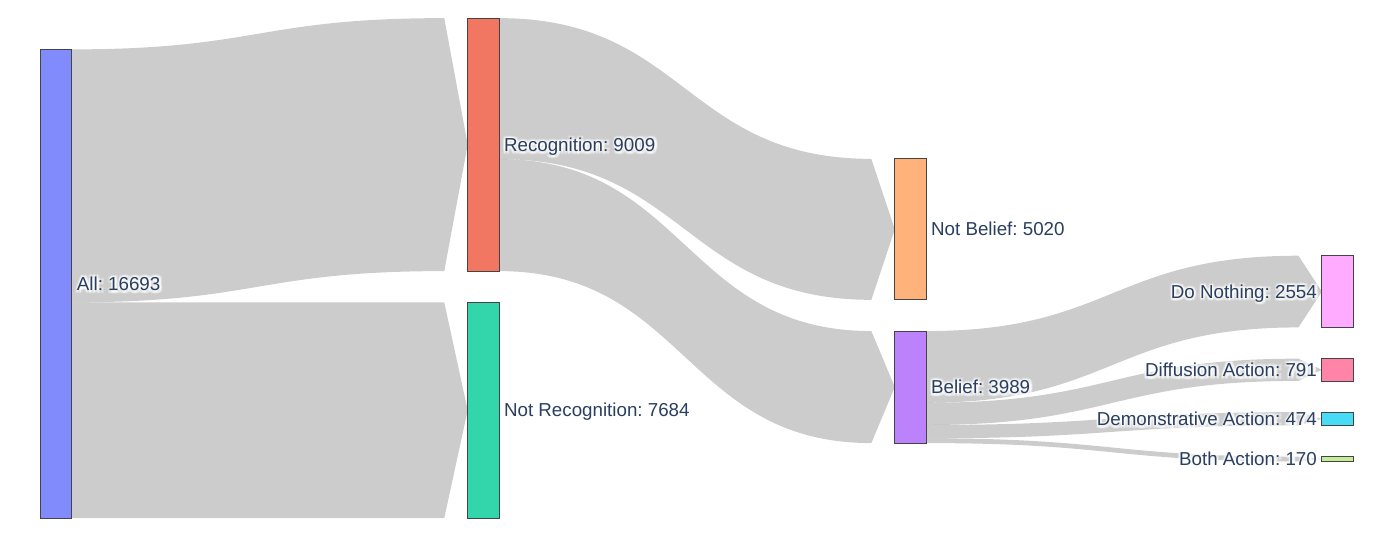}
        \caption{Sankey diagram showing the sequential stages of conspiracy theory engagement in Japan.}
    \end{subfigure}
    \caption{Sankey diagram showing the sequential stages of conspiracy theory engagement in Japan and the U.S. This figure visualizes the number of respondents who reported recognizing, believing, and acting upon at least one conspiracy theory. }
    \label{sankey_figure}
\end{figure}

Figure~\ref{sankey_figure} shows Sankey diagrams illustrating how respondents progress through the three key stages of conspiracy theory engagement: recognition, belief, and action.
In the U.S., 12,187 of the 13,578 respondents (approximately 90\%) recognized at least one conspiracy theory, and 8,415 of these (69\%) believed in at least one.
Among those believers, 4,341 individuals took demonstrative action, whereas 4,341 took diffusion action. 
In Japan, 9,009 of the 16,693 respondents (roughly 54\%) recognized at least one conspiracy theory, with 3,989 (24\%) believing in at least one.
Of these believers, 644 took demonstrative action and 961 participated in diffusion action.
These findings reveal two key patterns. 
First, a substantially larger share of U.S. respondents recognized and believed in conspiracy theories, leading to higher rates of active engagement. 
Second, in both countries, only a fraction of those who believe in conspiracy theories proceed to either demonstrative or diffusion action, suggesting that belief alone does not automatically lead to active engagement.
Sankey diagrams for each individual conspiracy theory are provided in Figure S1 and S2 in the Supplementary Information.
The specific types of Demonstrative Action and Diffusion Action associated with each conspiracy theory are detailed in Table S3 - S6 in the of Supplementary Information.

\begin{figure}[t]
    \centering
    \begin{subfigure}{0.45\linewidth}
        \centering
        \includegraphics[width=\linewidth]{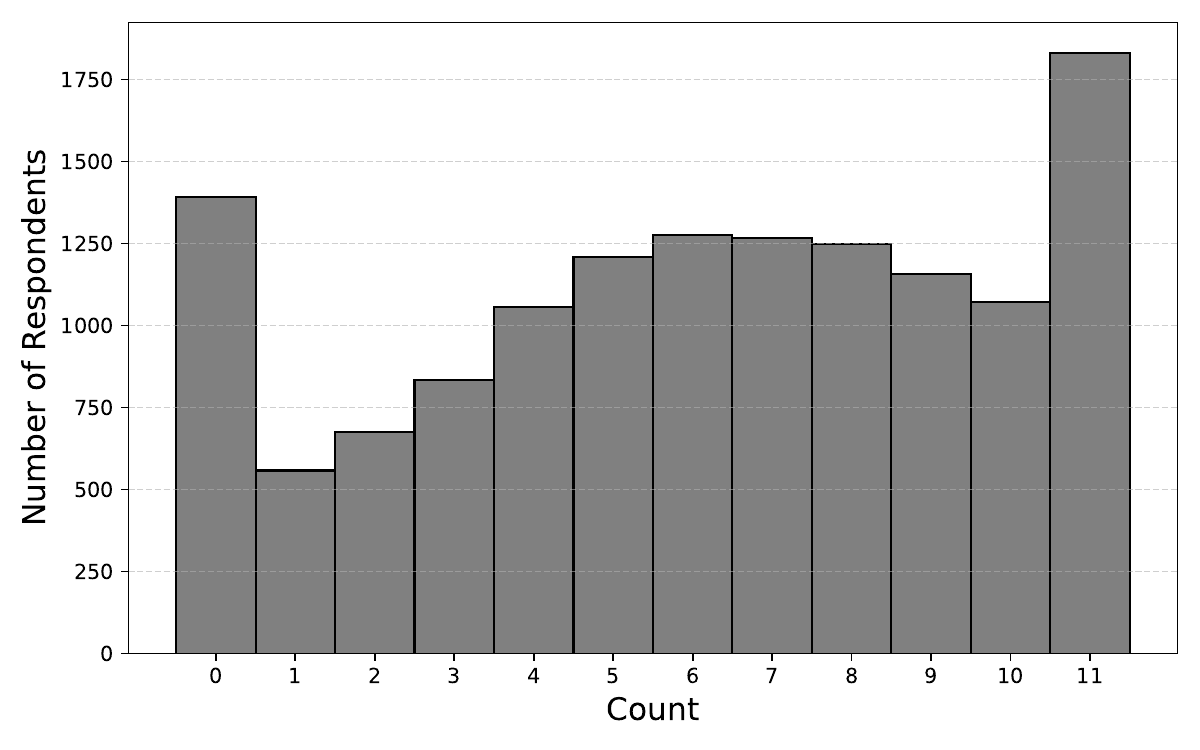}
        \caption{In the U.S.}
    \end{subfigure}
    \hfill
    \begin{subfigure}{0.45\linewidth}
        \centering
        \includegraphics[width=\linewidth]{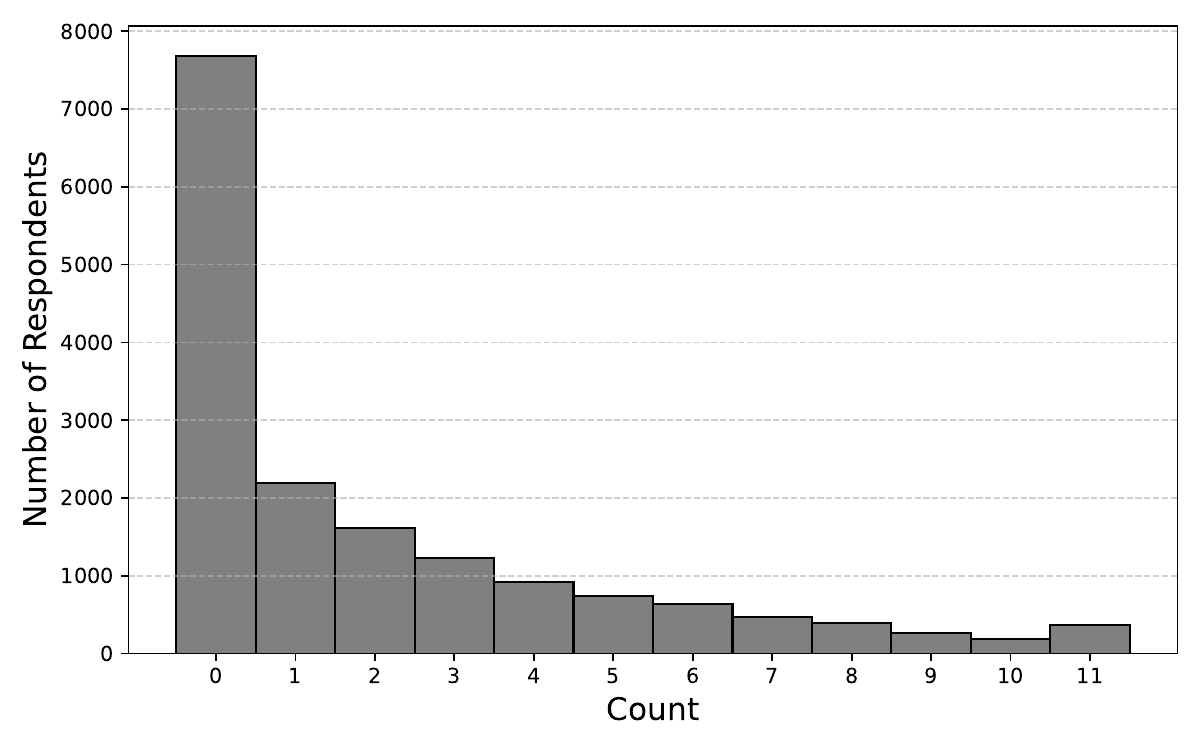}
        \caption{In Japan}
    \end{subfigure}
    \caption{Distribution of the number of conspiracy theories recognized by respondents in the U.S. and Japan. Respondents were asked whether they recognize each of the 11 conspiracy theories.}
    \label{recognition_histogram}
\end{figure}

\begin{figure}[t]
    \centering
    \begin{subfigure}{0.45\linewidth}
        \centering
        \includegraphics[width=\linewidth]{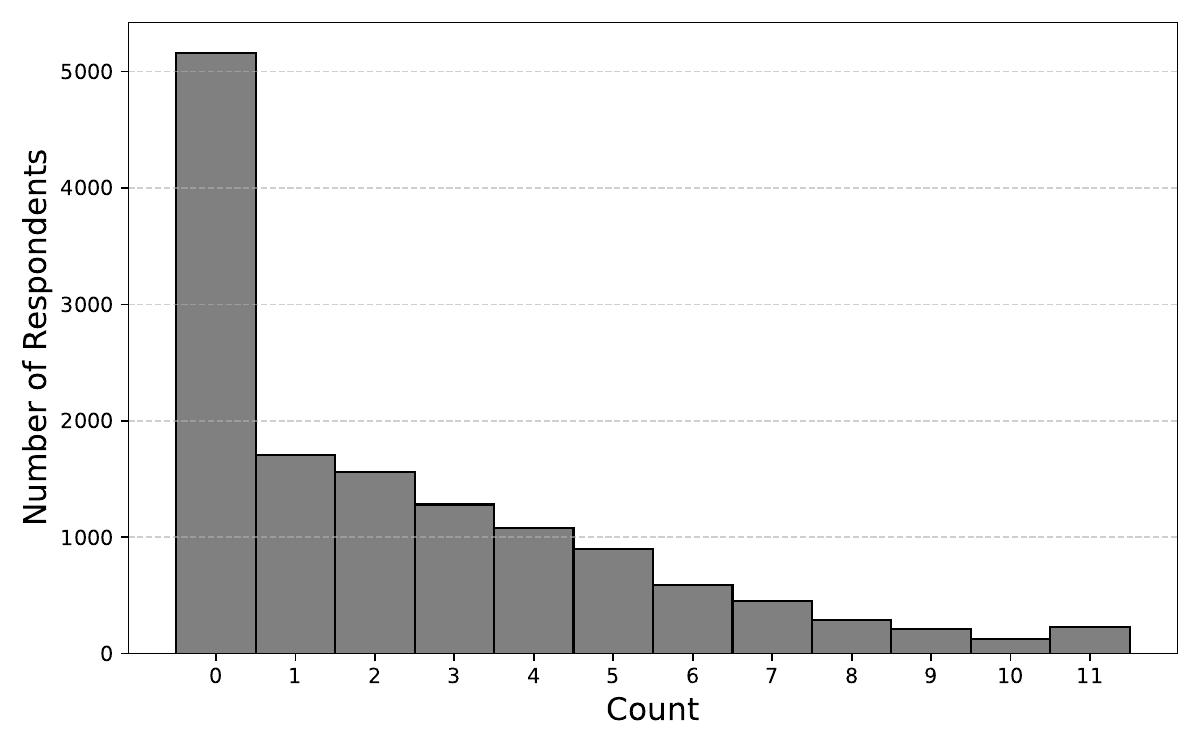}
        \caption{In the U.S.}
    \end{subfigure}
    \hfill
    \begin{subfigure}{0.45\linewidth}
        \centering
        \includegraphics[width=\linewidth]{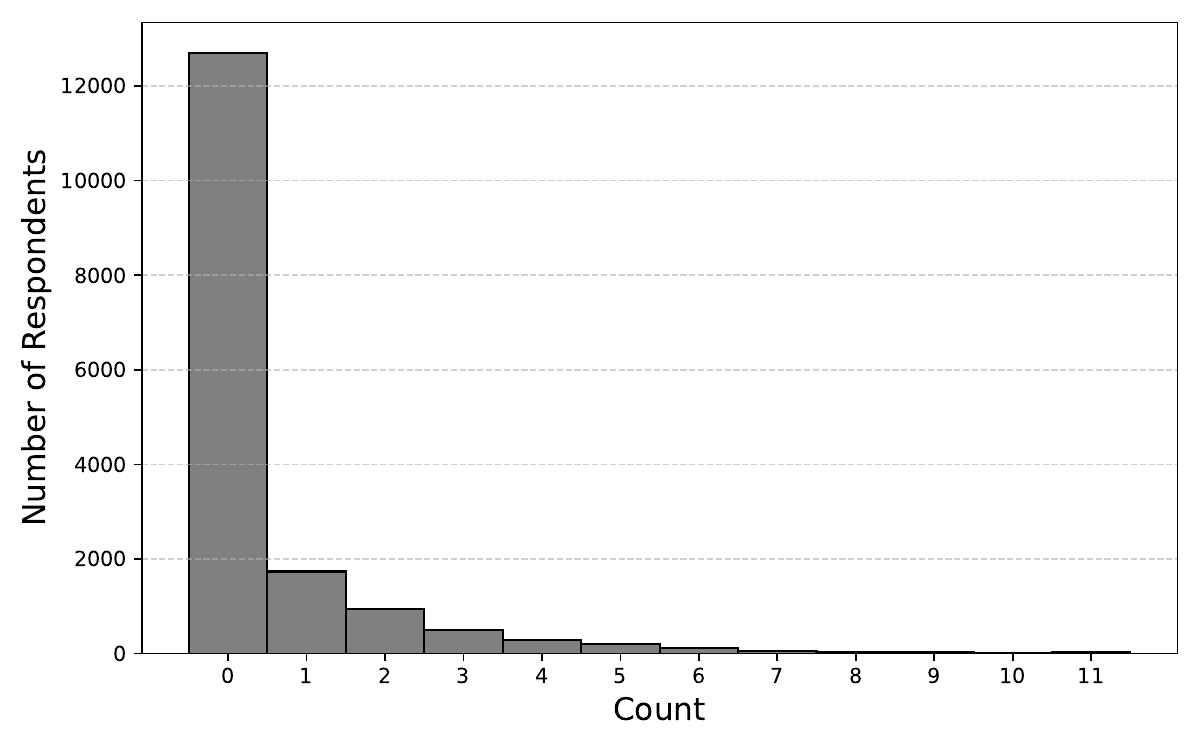}
        \caption{In Japan}
    \end{subfigure}
    \caption{Distribution of the number of conspiracy theories identified as belief by respondents in the U.S. and Japan. 
    Respondents were asked whether they believe each of the 11 conspiracy theories.}
    \label{belief_histogram}
\end{figure}

Figures~\ref{recognition_histogram} and \ref{belief_histogram} show the distributions of the number of conspiracy theories recognized and believed, by respondents in the U.S. and Japan, respectively.
In the U.S. (Fig.\ref{recognition_histogram} (a)), the distribution for recognition is skewed toward higher counts, indicating that many respondents are aware of multiple conspiracy theories.
By contrast, Japanese respondents (Fig.\ref{recognition_histogram} (b)) tend to recognize none or only a small number of the listed theories.
A similar pattern emerges in the belief distributions (Fig.\ref{belief_histogram}). 
In the U.S. (Fig.\ref{belief_histogram} (a)), a substantial proportion of respondents report believing in at least one conspiracy theory.
Meanwhile, in Japan (Fig.\ref{belief_histogram} (b),) most respondents either believe in none or only one or two theories. 
These overall patterns are consistent with the Sankey diagrams shown in Figure~\ref{sankey_figure}, demonstrating that American respondents not only recognize conspiracies at higher rates but also tend to believe in a larger range of them, compared to their Japanese counterparts.

\begin{figure}[t]
    \centering

    \begin{subfigure}{0.47\linewidth}
        \centering
        \includegraphics[width=\linewidth]{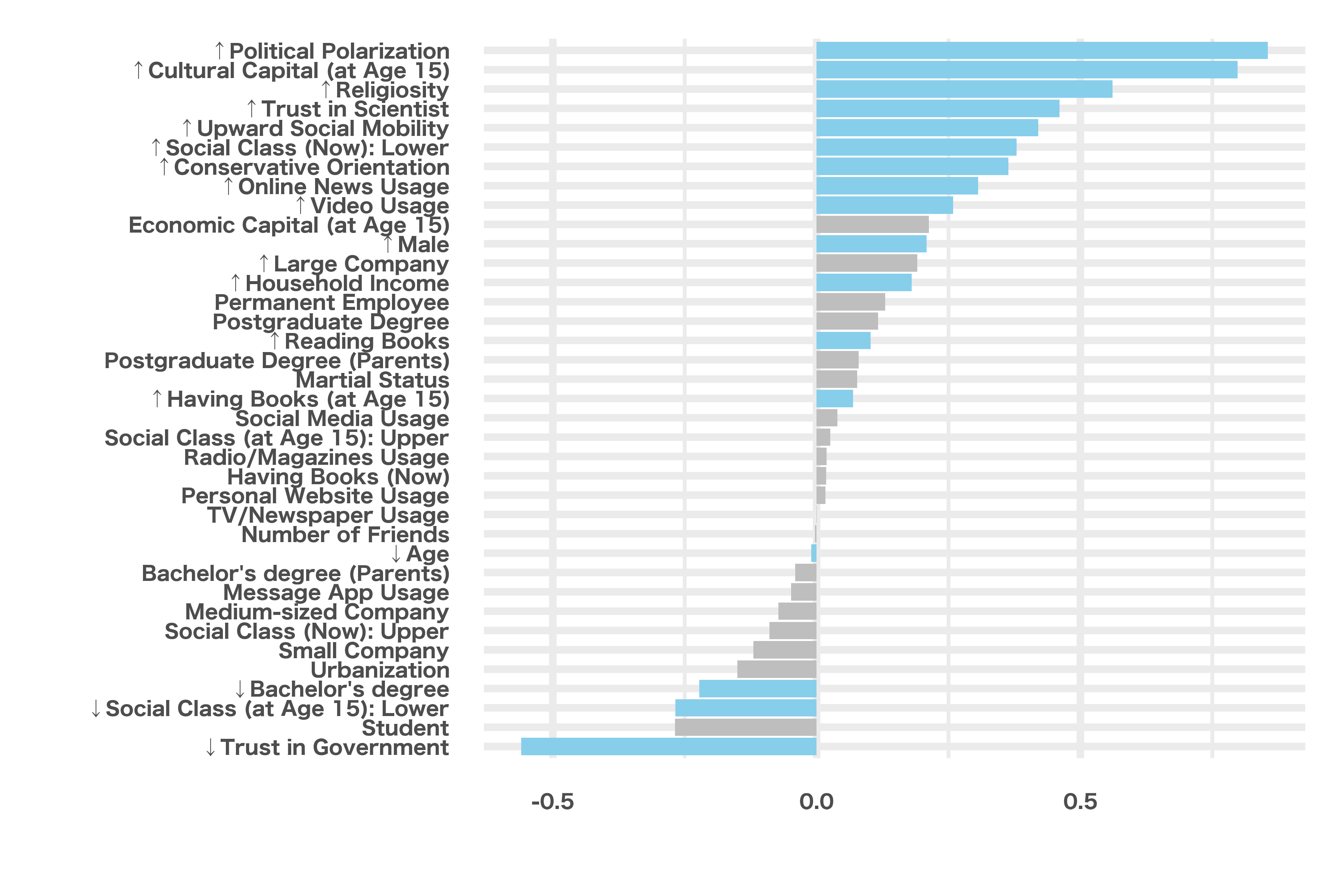}
        \caption{Effects of explanatory variables on conspiracy theory \textbf{Recognition} in the U.S.}
    \end{subfigure}
    \hfill
    \begin{subfigure}{0.47\linewidth}
        \centering
        \includegraphics[width=\linewidth]{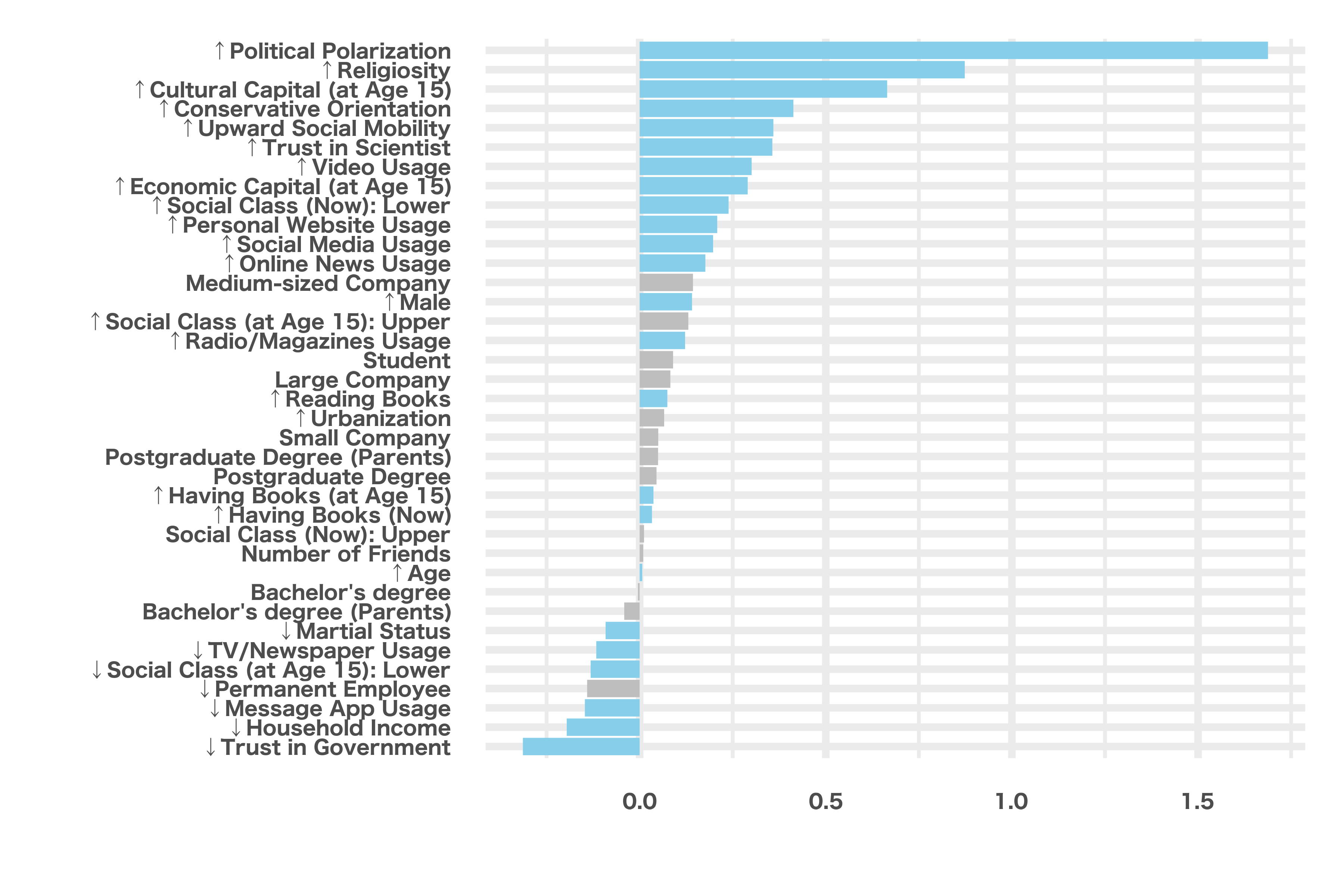}
        \caption{Effects of explanatory variables on conspiracy theory \textbf{Recognition} in Japan.}
    \end{subfigure}

    \begin{subfigure}{0.47\linewidth}
        \centering
        \includegraphics[width=\linewidth]{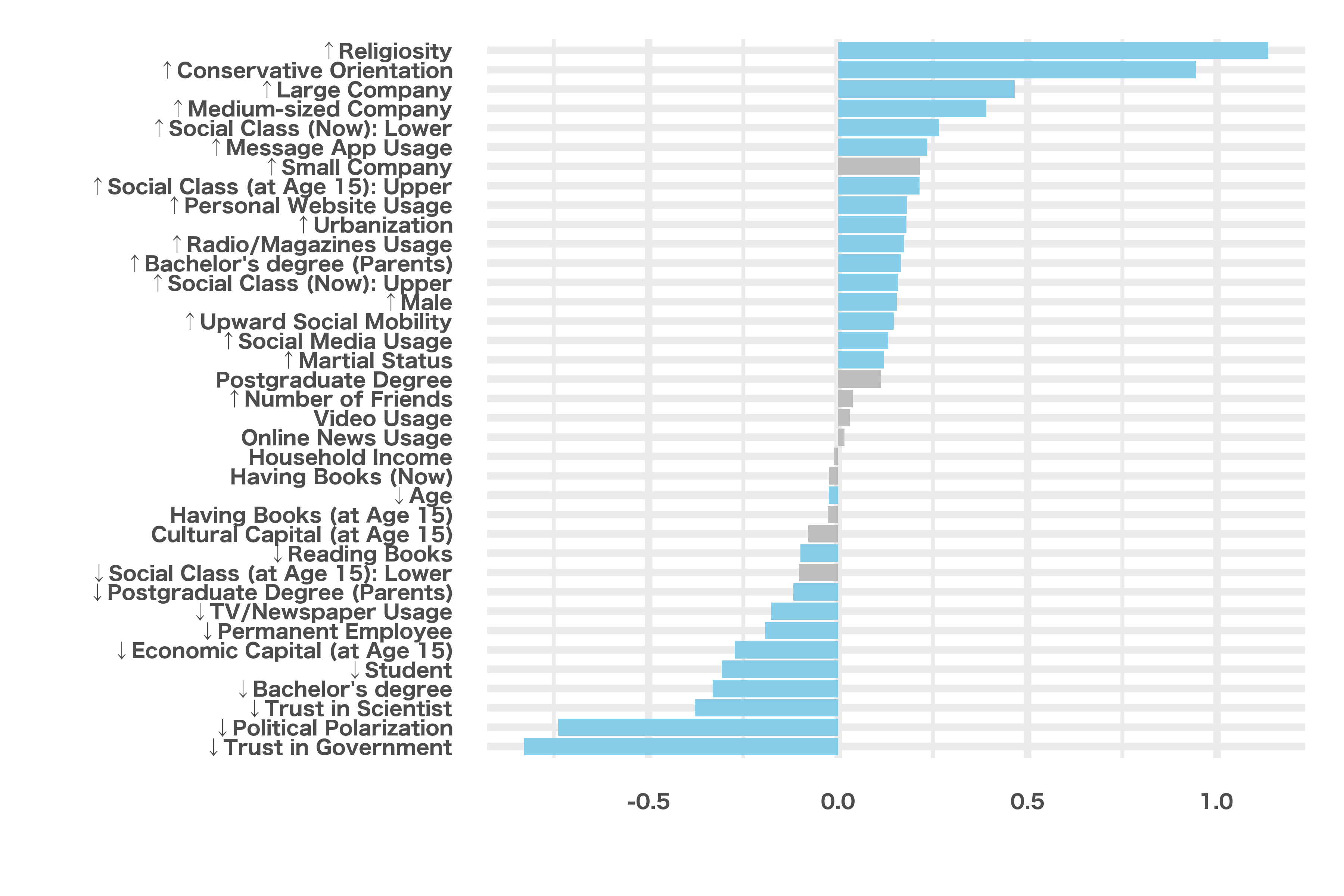}
        \caption{Effects of explanatory variables on conspiracy theory \textbf{Belief} in the U.S.}
    \end{subfigure}
    \hfill
    \begin{subfigure}{0.47\linewidth}
        \centering
        \includegraphics[width=\linewidth]{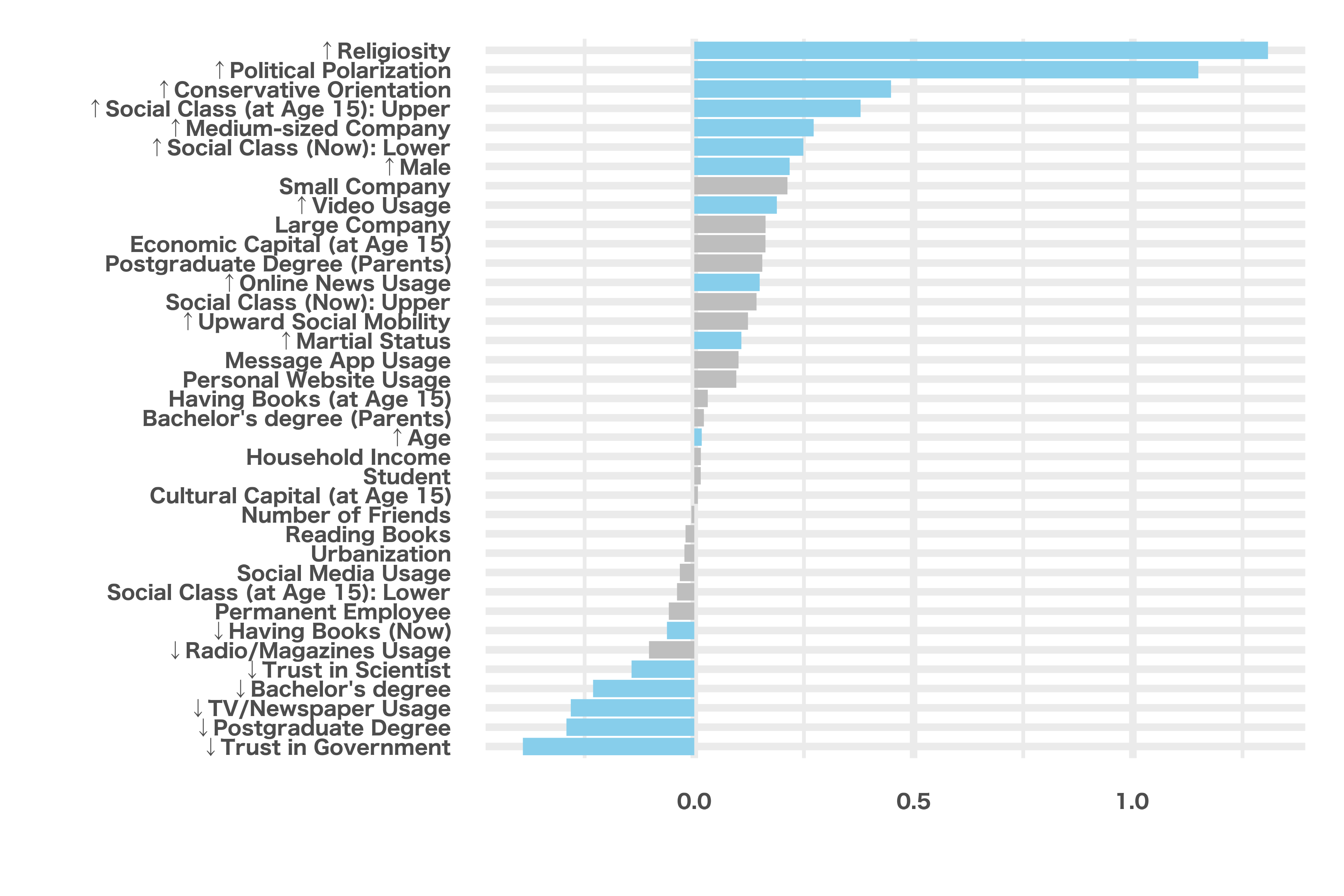}
        \caption{Effects of explanatory variables on conspiracy theory \textbf{Belief} in Japan.}
    \end{subfigure}

    \begin{subfigure}{0.47\linewidth}
        \centering
        \includegraphics[width=\linewidth]{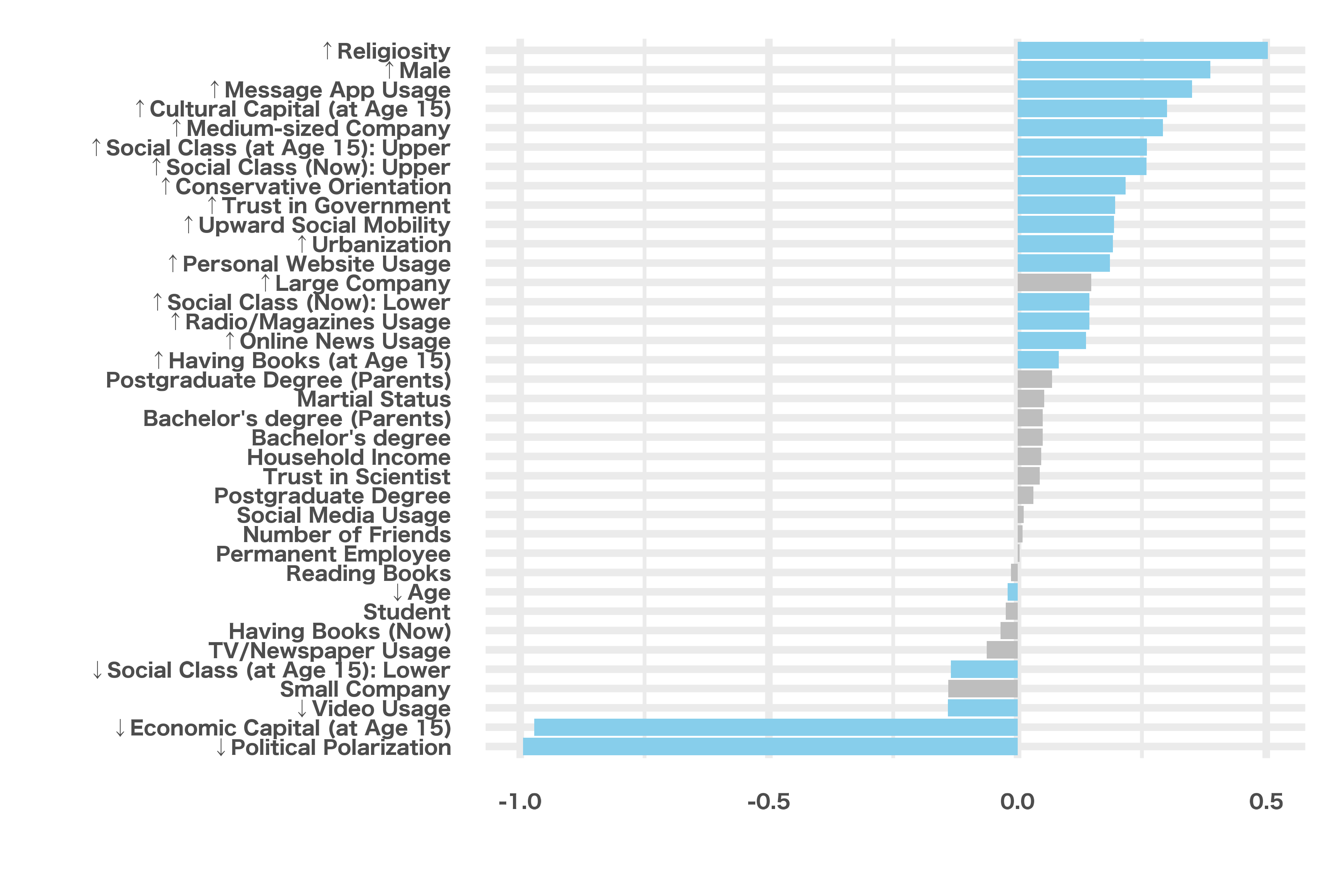}
        \caption{Effects of explanatory variables on conspiracy theory \textbf{Demonstrative Action} in the U.S.}
    \end{subfigure}
    \hfill
    \begin{subfigure}{0.47\linewidth}
        \centering
        \includegraphics[width=\linewidth]{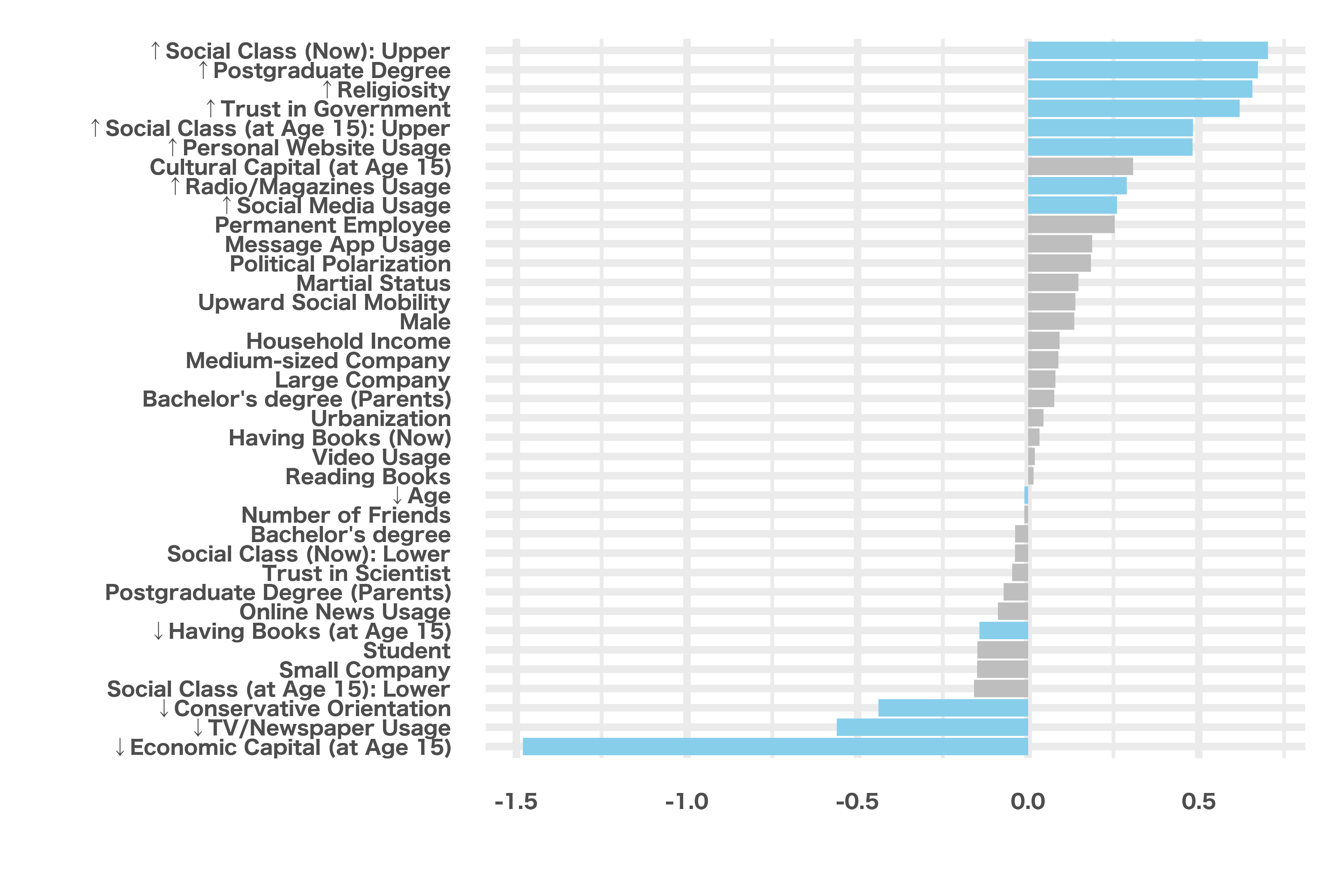}
        \caption{Effects of explanatory variables on conspiracy theory \textbf{Demonstrative Action} in Japan.}
    \end{subfigure}

    \begin{subfigure}{0.47\linewidth}
        \centering
        \includegraphics[width=\linewidth]{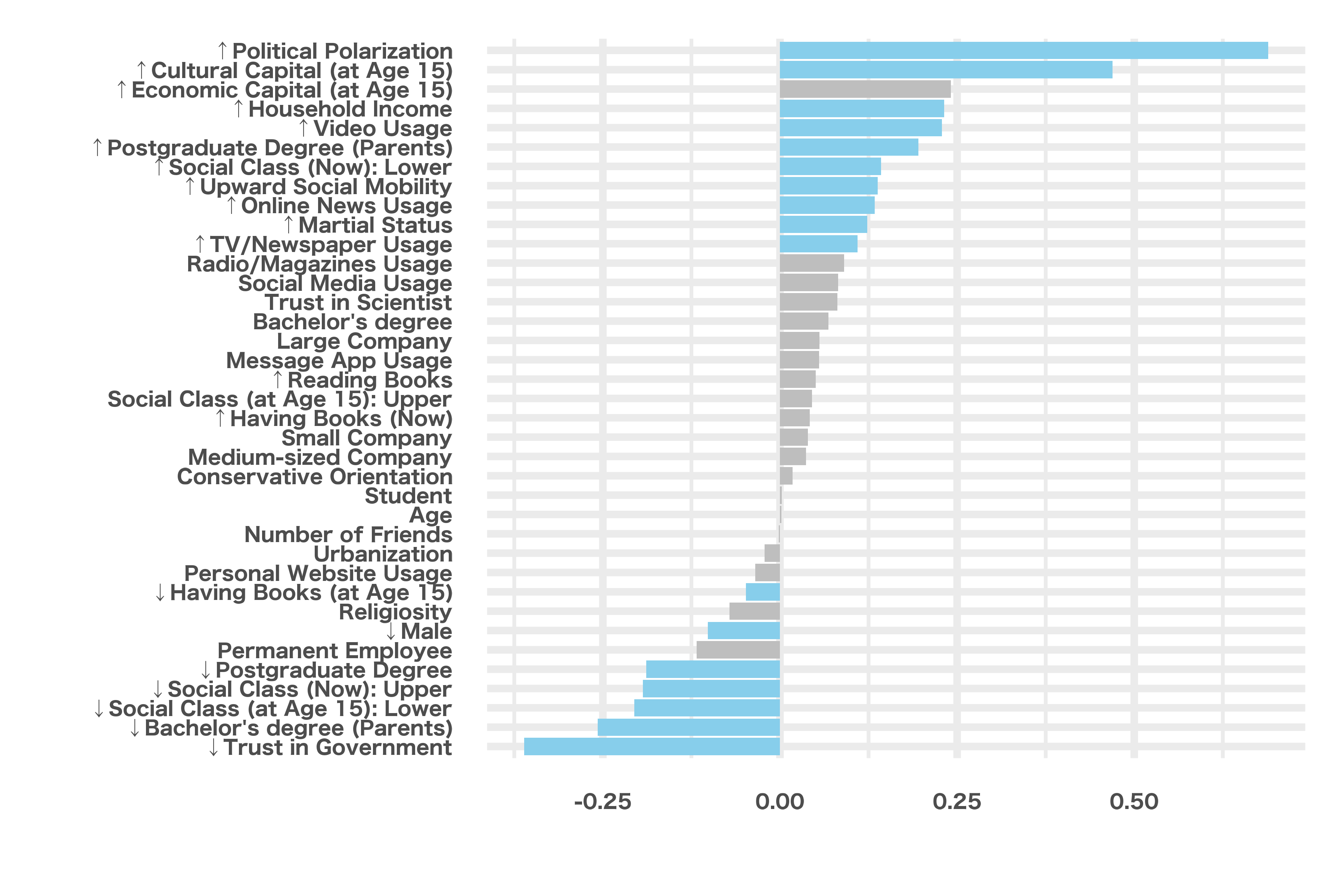}
        \caption{Effects of explanatory variables on conspiracy theory \textbf{Diffusion Action} in the U.S.}
    \end{subfigure}
    \hfill
    \begin{subfigure}{0.47\linewidth}
        \centering
        \includegraphics[width=\linewidth]{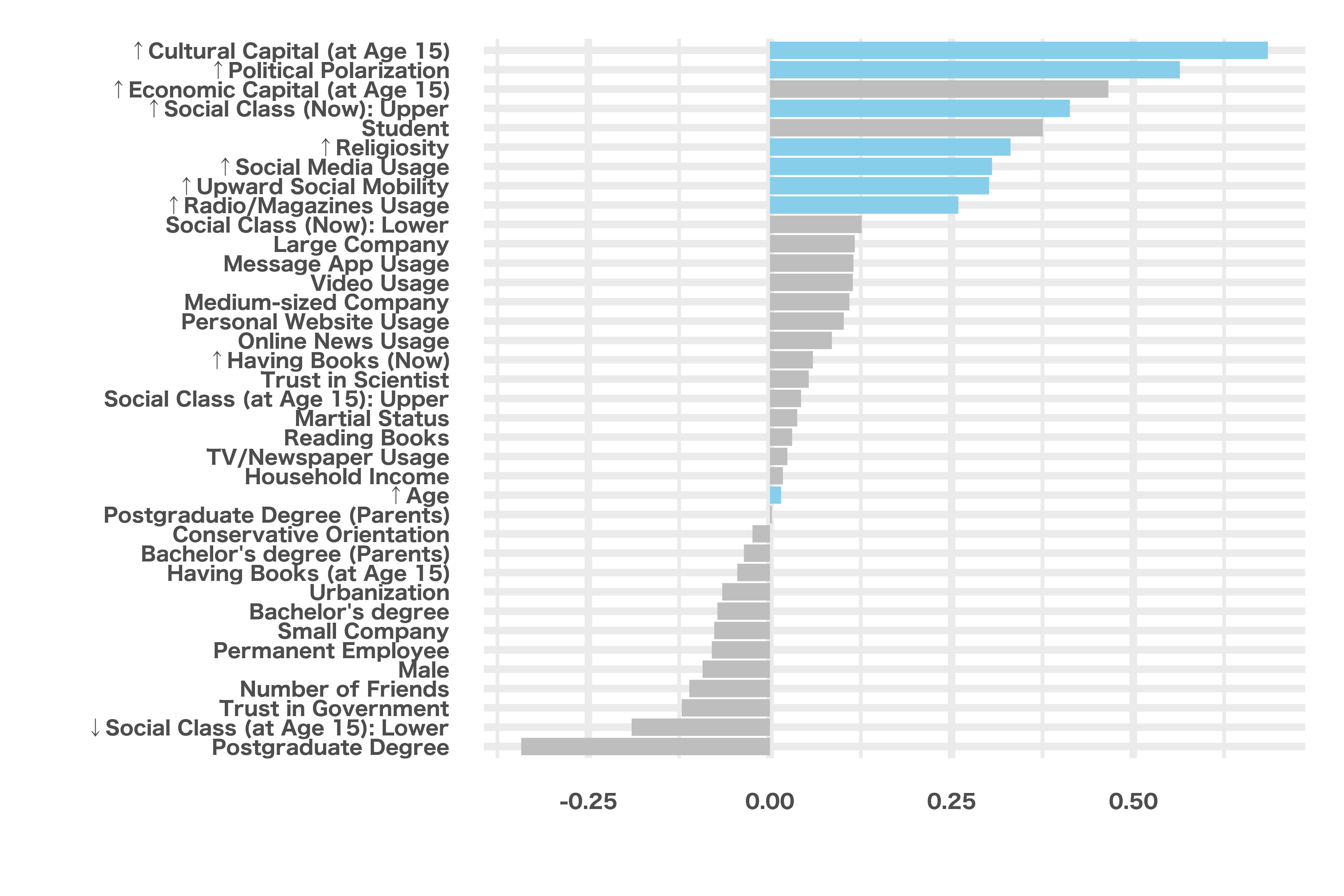}
        \caption{Effects of explanatory variables on conspiracy theory \textbf{Diffusion Action} in \\ Japan.}
    \end{subfigure}
    \caption{
    Estimated effects of explanatory variables on each stage of conspiracy theory engagement (Recognition, Belief, Demonstrative Action, and Diffusion Action) in Japan and the U.S. 
    Each bar represents the estimated regression coefficient for an explanatory variable:
    \textbf{↑} indicates a statistically significant positive effect (P($\beta > 0$) > 0.95), while \textbf{↓} indicates a statistically significant negative effect (P($\beta < 0$) > 0.95). 
    Bars shown in skyblue denote significant effects based on 95\% credible intervals, while those shown in gray represent non-significant effects where the credible interval crosses zero.
    }
    \label{result_regression}
\end{figure}

\subsection*{Determinants of Recognition, Belief, and Action}
To systematically investigate the factors influencing conspiracy theory engagement, we applied a Bayesian Hierarchical Bernoulli model, which enables us to estimate the effects of key demographic, political, and social variables at each stage, Recognition, Belief, and Action, while accounting for both direct and indirect effects across stages.
This approach enables a comprehensive understanding of how individual characteristics and societal factors contribute to the progression from mere awareness of conspiracy theories to active dissemination or demonstration.

The results of our analysis are summarized in Figure~\ref{result_regression}, which presents the estimated effects of explanatory variables on each stage of conspiracy theory engagement in both the U.S. and Japan. 
To ensure the robustness of our findings, detailed convergence diagnostics and exact parameter estimates are provided in Tables S7 - S14 in the Supplementary Information.
Additionally, a more granular analysis focusing on individual conspiracy theories is available in Figures S9 - S19 in the Supplementary Information for the U.S. and Figures S20 - S30 in the Supplementary Information for Japan.
The following section examines the specific determinants at each stage and identifies overarching patterns across both countries. 
By comparing similarities and differences in the factors influencing each stage, we highlight the underlying mechanisms that shape conspiracy theory engagement and the extent to which they are driven by universal or context-dependent influences.

\subsubsection*{Recognition in Conspiracy Theory}








The recognition of conspiracy theories is significantly influenced by political orientation, social capital, media habit patterns, and demographic attributes. 
Political ideology plays a central role in shaping exposure to conspiracy narratives.
\textbf{Conservative Orientation} (U.S.: OR = $\exp(0.363) = 1.44$, Japan: OR = $\exp(0.413) = 1.51$) and \textbf{Political Polarization} (U.S.: OR = $\exp(0.855) = 2.35$, Japan: OR = $\exp(1.688) = 5.41$) exhibit strong positive associations with conspiracy theory recognition. 
These findings align with existing literature suggesting that politically polarized individuals are more likely to question mainstream narratives and engage with alternative sources of information, including conspiracy theories~\cite{vanKrouwel_2022,littrell2023knowingly}. 
The influence of conservatism may also reflect the ideological framing of specific conspiracy theories included in our study. 
For instance, U.consp6 ``Democratic Party members are involved in organized criminal activities''  and J.consp4 ``Foreign residents in Japan are manipulating politicians and the media'' are narratives that resonate more strongly with conservative worldviews, potentially amplifying their recognition rates. 
However, the consistent association across both countries suggests a broader tendency for politically conservative and polarized individuals to engage with information about conspiracy thory.

Beyond political factors, cultural capital plays a critical role in exposure to conspiracy theories. 
Indicators such as \textbf{Cultural Capital at Age 15} (U.S.: OR = $\exp(0.798) = 2.22$, Japan: OR = $\exp(0.665) = 1.94$), \textbf{Having Books at Age 15} (U.S.: OR = $\exp(0.069) = 1.07$, Japan: OR = $\exp(0.037) = 1.04$), and \textbf{Reading Books} (U.S.: OR = $\exp(0.103) = 1.11$, Japan: OR = $\exp(0.074) = 1.08$) are positively associated with recognition. 
This suggests that individuals who had access to diverse information sources from an early age are more likely to encounter and recognize conspiracy theories. 
Regular reading habits may encourage active information-seeking behavior, increasing exposure to both mainstream and fringe narratives.

Media usage patterns further shape the likelihood of conspiracy theory recognition. 
\textbf{Online News Usage} (\text{U.S.}: OR = $\exp(0.306) = 1.36$, Japan: OR = $\exp(0.177) = 1.19$) and \textbf{Video Usage} (\text{U.S.}: OR = $\exp(0.258) = 1.29$, Japan: OR = $\exp(0.301) = 1.35$) positively correlate with recognition, consistent with prior research linking YouTube and online platforms to conspiracy theory exposure~\cite{ha2022conspiracy,ginossar2022cross}.  
Algorithm-driven recommendations and user engagement mechanisms on these platforms may increase encounters with conspiracy-related content.

Several additional demographic and attitudinal factors are linked to conspiracy recognition. 
\textbf{Male gender} (\text{U.S.}: OR = $\exp(0.209) = 1.23$, Japan: OR = $\exp(0.141) = 1.15$), \textbf{Religiosity} (\text{U.S.}: OR = $\exp(0.561) = 1.75$, Japan: OR = $\exp(0.873) = 2.39$), \textbf{Lower Social Class (Now)} (\text{U.S.}: OR = $\exp(0.379) = 1.46$, Japan: OR = $\exp(0.239) = 1.27$), and \textbf{Trust in Scientists} (\text{U.S.}: OR = $\exp(0.460) = 1.58$, Japan: OR = $\exp(0.356) = 1.43$) are all positively associated with recognition. 
A particularly striking finding is the strong negative effect of \textbf{Trust in Government} (U.S.: OR = $\exp(-0.560) = 0.57$, Japan: OR = $\exp(-0.314) = 0.73$). 
This pattern aligns with increasing political distrust, where partisan divides and media fragmentation have fueled skepticism toward government institutions~\cite{pierre2020mistrust}.
These patterns suggest that individuals who engage with structured belief systems, whether religious, ideological, or scientific, may have higher exposure to conspiracy-related narratives.

While these broad trends hold across both countries, several notable differences emerge. 
\textbf{Household Income} and \textbf{Permanent Employment} increase recognition in the U.S. (\text{OR} = $\exp(0.180) = 1.20$, $\exp(0.130) = 1.14$) but decrease recognition in Japan (\text{OR} = $\exp(-0.196) = 0.82$, $\exp(-0.141) = 0.87$). 
This divergence may reflect differences in how socioeconomic status influences information consumption. 
In the U.S., higher-income individuals may have greater access to diverse media, including alternative news and online communities where conspiracy theories circulate~\cite{schulte2022income}. 
In contrast, in Japan, economic stability and full-time employment may reduce the incentive to seek non-mainstream narratives.

Age effects also differ across countries.
In the U.S., younger individuals are more likely to recognize conspiracy theories (\text{OR} = $\exp(-0.010) = 0.99$ per year increase), whereas in Japan, older individuals exhibit higher recognition (\text{OR} = $\exp(0.007) = 1.01$ per year increase). 
This divergence may reflect generational differences in media consumption and digital exposure.
A similar divergence is observed in the relationship between traditional media usage and recognition.
In the U.S., reliance on \textbf{TV/Newspaper Usage} has no influence on recognition (\text{OR} = $\exp(0.001) = 1.00$), whereas in Japan, it is significantly associated with lower recognition (\text{OR} = $\exp(-0.116) = 0.89$). 
This pattern likely stems from structural differences in national media ecosystems. 
In the U.S., mainstream media frequently report on conspiracy theories, even in critical contexts, increasing public exposure.
In contrast, Japanese mainstream media tend to avoid direct coverage of conspiracy theories, reducing their visibility among audiences who primarily consume traditional news sources.

\begin{figure}[t]
    \centering
    \begin{subfigure}{0.45\linewidth}
        \centering
        \includegraphics[width=\linewidth]{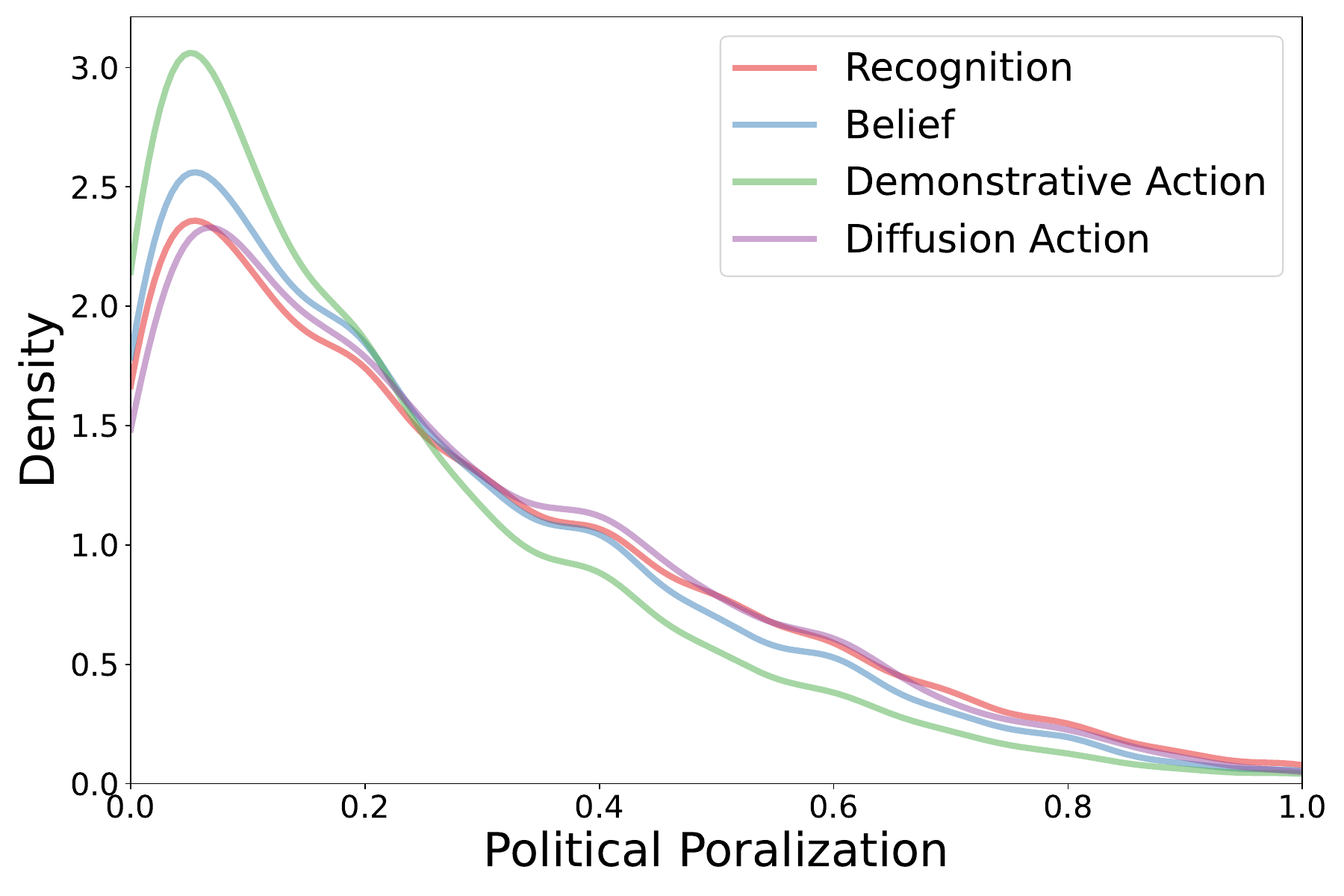}
        \caption{Distribution of Political Polarization Scores in the U.S.}
    \end{subfigure}
    \hfill
    \begin{subfigure}{0.45\linewidth}
        \centering
        \includegraphics[width=\linewidth]{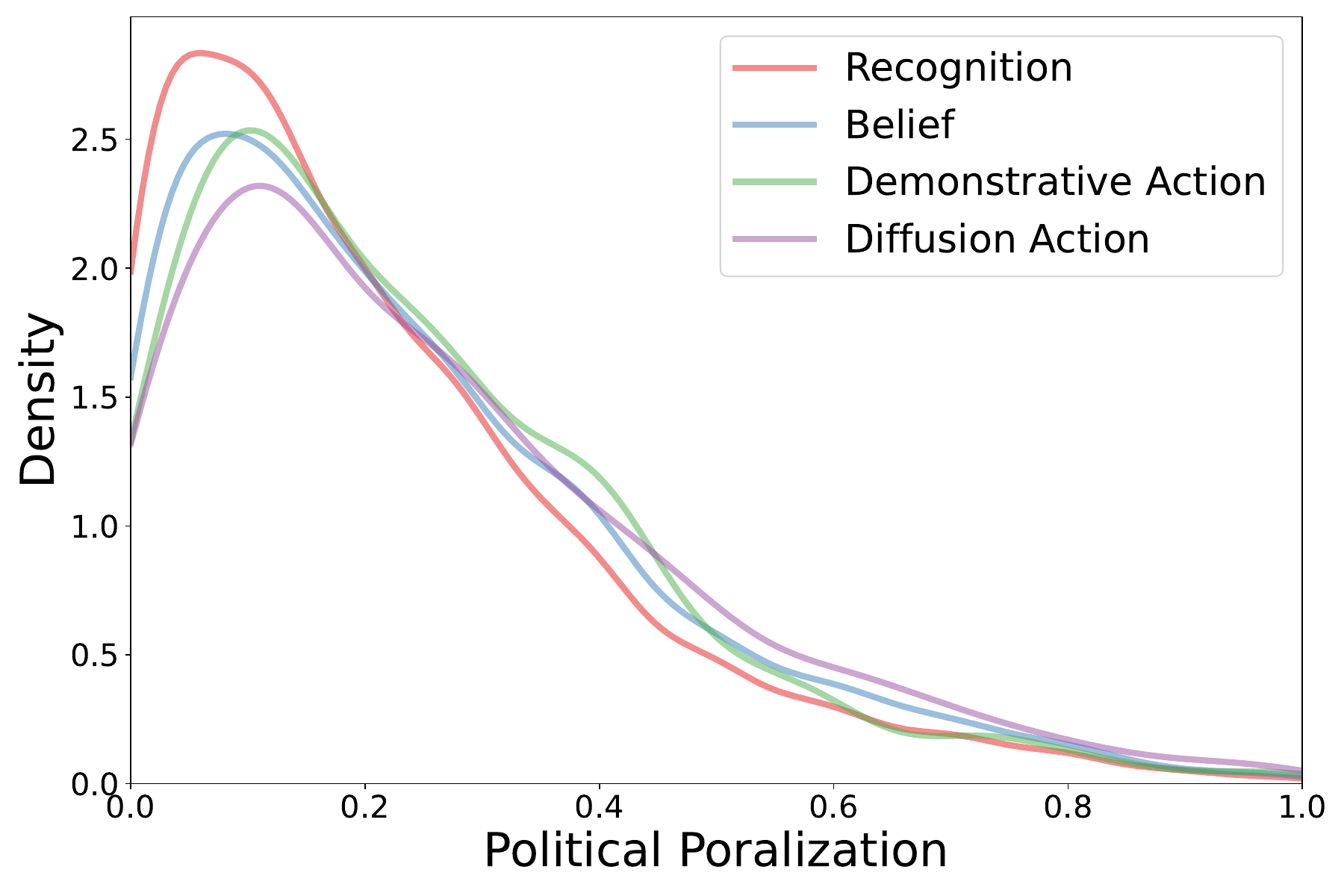}
        \caption{Distribution of Political Polarization Scores in Japan.}
    \end{subfigure}
    \caption{The distribution of political polarization scores for users across each stage: Recognition, Belief, Demonstrative Action, and Diffusion Action. }
    \label{political_poralization}
\end{figure}

\subsubsection*{Belief in Conspiracy Theory}
As discussed in \nameref{background}, numerous existing studies have examined the relationship between conspiracy beliefs and individual factors. and our findings largely align with and support these prior results.
Similar to recognition, political ideology strongly influences conspiracy belief. 
\textbf{Conservative Orientation} is positively associated with belief in both countries (U.S.: OR = $\exp(0.945) = 2.57$, Japan: OR = $\exp(0.449) = 1.57$), which align with previous research~\cite{miller2016conspiracy, djordjevic2021beyond, grzesiak2009right, van2021paranoid}. 
Furthermore, \textbf{Religiosity} exhibits a particularly strong association with belief (U.S.: OR = $\exp(1.135) = 3.11$, Japan: OR = $\exp(1.308) = 3.70$), reinforcing previous findings that religious belief systems and conspiratorial worldviews often share structural similarities in explaining uncertainty and unseen threats~\cite{galliford2017individual, lowicki2022does, lobato2014examining,frenken2023relation}. 

Trust in authoritative institutions functions as a crucial determinant of belief.
Both \textbf{Trust in Scientists} (U.S.: OR = $\exp(-0.379) = 0.68$, Japan: OR = $\exp(-0.143) = 0.87$) and \textbf{Trust in Government} (U.S.: OR = $\exp(-0.828) = 0.44$, Japan: OR = $\exp(-0.391) = 0.68$) exhibit significant negative associations with belief. 
These results align with research suggesting that individuals who distrust official institutions are more inclined to adopt alternative narratives, including conspiracy theories~\cite{jolley2020pylons, edelson2017effect, nyhan2017media, kofta2020breeds, vranic2022did, van2020engagement, brzezinski2020belief, roozenbeek2020susceptibility}.

The type of media usage habit further shapes their likelihood of believing in conspiracy theories. 
\textbf{Personal Website Usage} increases belief in both countries (U.S.: OR = $\exp(0.183) = 1.20$, Japan: OR = $\exp(0.096) = 1.10$), indicating that alternative, user-curated online spaces contribute to the reinforcement of conspiratorial thinking. 
\textbf{Message App Usage} also exhibits a positive correlation with belief in both countries (U.S.: OR = $\exp(0.235) = 1.26$, Japan: OR = $\exp(0.101) = 1.11$), which highlights the role of interpersonal digital communication channels in facilitating the conspiracy beliefs.
By contrast, engagement with traditional media such as \textbf{TV/Newspaper Usage} is negatively correlated with belief (U.S.: OR = $\exp(-0.177) = 0.84$, Japan: OR = $\exp(-0.282) = 0.75$).

Social and economic status also play a role in belief formation. 
Individuals who perceive themselves as belonging to \textbf{Social Class (Now): Lower} are more likely to endorse conspiracy theories in both countries (U.S.: OR = $\exp(0.266) = 1.30$, Japan: OR = $\exp(0.249) = 1.28$), consistent with prior studies linking economic precarity to conspiratorial thinking~\cite{casara2022impact, jetten2022economic}.
At the same time, \textbf{Upward social mobility} is positively correlated with belief (U.S.: OR = $\exp(0.147) = 1.16$, Japan: OR = $\exp(0.122) = 1.13$), suggesting that not only economic hardship but also experiences of social transition may heighten receptivity to conspiracy beliefs.

Additional demographic and attitudinal factors are linked to conspiracy belief; \textbf{Male gender} (U.S.: OR = $\exp(0.155) = 1.17$, Japan: OR = $\exp(0.217) = 1.24$), \textbf{Marital Status} (U.S.: OR = $\exp(0.121) = 1.13$, Japan: OR = $\exp(0.107) = 1.11$), and \textbf{Social Class (at Age 15): Upper} (U.S.: OR = $\exp(0.214) = 1.24$, Japan: OR = $\exp(0.379) = 1.46$), whereas \textbf{Bachelor’s Degree} (U.S.: OR = $\exp(-0.331) = 0.72$, Japan: OR = $\exp(-0.231) = 0.79$) exhibits a negative association.

Despite these broad similarities, cross-national differences emerge in the demographic and behavioral predictors of conspiracy belief.
One notable distinction is the effect of \textbf{Age}, which negatively correlates with belief in the U.S. (OR = $\exp(-0.025) = 0.98$ per year increase) but positively correlates in Japan (OR = $\exp(0.017) = 1.02$ per year increase). 
This suggests that in the U.S., younger individuals are more susceptible to conspiracy beliefs, whereas in Japan, older individuals demonstrate higher belief levels. 
These opposing trends may reflect generational differences in media consumption habits and exposure to digital misinformation ecosystems. 
In Japan, younger cohorts have been the primary target of recent information literacy initiatives, potentially mitigating their exposure to conspiracy-related content, while middle-aged and older populations might not have benefited from such initiatives.
Educational attainment also plays a more prominent role in Japan, where holding a \textbf{postgraduate degree} is negatively associated with belief (OR = $\exp(-0.291) = 0.75$), whereas in the U.S., postgraduate education shows little effect (OR = $\exp(0.112) = 1.12$).

Differences in media usage habit further illustrate cultural contrasts in belief formation.
In the U.S., belief is positively associated with \textbf{Social Media Usage} (OR = $\exp(0.132) = 1.14$) and \textbf{Radio/Magazine Usage} (OR = $\exp(0.174) = 1.19$), while in Japan, both factors show weak negative associations (social media: OR = $\exp(-0.033) = 0.97$; radio/magazines: OR = $\exp(-0.103) = 0.90$). 
Conversely, \textbf{Video Usage} plays a stronger role in Japan (OR = $\exp(0.188) = 1.21$) than in the U.S. (OR = $\exp(0.031) = 1.03$), suggesting that the influence of video-based conspiracy may vary by country. 
Additionally, \textbf{Online News Usage} has little effect in the U.S. (OR = $\exp(0.017) = 1.02$) but is negatively associated with belief in Japan (OR = $\exp(-0.149) = 0.86$).
The extent to which digital news platforms contribute to or counteract conspiracy beliefs likely depends on differences in the credibility and ideological orientation of each online news source.

One of the most striking findings is the contrasting effect of \textbf{Political Polarization} on conspiracy belief. 
While existing studies consistently indicate that individuals with extreme political polarization tend to exhibit stronger conspiracy beliefs~\cite{sutton2020conspiracy, imhoff2022conspiracy, van2015political, nera2021power, krouwel2017does}, our results reveal an opposite effect between the two countries. 
In the U.S., political polarization is negatively associated with belief (OR = $\exp(-0.739) = 0.48$), whereas in Japan, it is positively correlated (OR = $\exp(1.149) = 3.16$). 
This finding is particularly surprising for the U.S., where higher polarization scores are associated with lower conspiracy belief, contradicting previous research. 
To further explore this dynamic, Figure~\ref{political_poralization} presents the distribution of political polarization scores across different engagement stages. 
Notably, in the U.S., individuals with extreme polarization scores exhibit higher recognition rates than belief rates, indicating that politically polarized individuals encounter conspiracy theories but are not necessarily convinced by them. 
In contrast, in Japan, individuals who believe in conspiracy theories tend to have higher polarization scores than those who merely recognize them. 
This divergence highlights the importance of distinguishing between different stages of conspiracy engagement. 
Previous research has predominantly focused on belief while neglecting the influence of recognition, potentially leading to an overestimation of the effect of political polarization. 
By accounting for the sequential nature of conspiracy theory exposure and adoption, our model suggests that in the U.S., political polarization primarily influences recognition rather than belief. 
These findings highlight the necessity of incorporating multi-stage frameworks in conspiracy theory research to capture the nuanced interactions between political ideology and belief formation.

\subsubsection*{Demonstrative Action in Conspiracy Theory}
In both countries, \textbf{Religiosity} positively correlates with the likelihood of demonstrative action (U.S.: OR = $\exp(0.504) = 1.66$, Japan: OR = $\exp(0.657) = 1.93$), mirroring similar patterns observed at earlier stages. 
\textbf{Age} exhibits a negative effects (U.S.: OR = $\exp(-0.020) = 0.98$, Japan: OR = $\exp(-0.011) = 0.99$), aligning with prior findings that younger individuals are more likely to participate in political demonstrations~\cite{PewResearch2019,Rengo2021,kwak2022measuring}. 
This tendency may be attributed to multiple factors, including physical capacity, and flexibility in time commitment. 
Additionally, younger individuals are often more active in online communities, which can facilitate mobilization and lower the barriers to collective action.
A striking departure from the pattern at earlier stages concerns \textbf{Trust in Government}, which here shows a positive association (U.S.: OR = $\exp(0.197) = 1.22$, Japan: OR = $\exp(0.620) = 1.86$).
These results suggest that, rather than driving protest through a sense of disillusionment or distrust in authorities, individuals who maintain faith in governmental institutions may become motivated to demand action or reform once they embrace conspiratorial narratives.
In both nations, certain media habits also relates to the demonstrative action; \textbf{Radio/Magazines Usage} (U.S.: OR = $\exp(0.144) = 1.15$, Japan: OR = $\exp(0.289) = 1.34$) and \textbf{Personal Website Usage} (U.S.: OR = $\exp(0.185) = 1.20$, Japan: OR = $\exp(0.482) = 1.62$) display consistently positive effects, whereas \textbf{TV/Newspaper Usage} lowers the probability of demonstrative action (U.S.: OR = $\exp(-0.062) = 0.94$, Japan: OR = $\exp(-0.561) = 0.57$).

Economic background and social class further shape the likelihood of demonstrative action.
Individuals from wealthier childhoods are less inclined to participate (\textbf{Economic Capital at Age 15}: U.S.: OR = $\exp(-0.972) = 0.38$, Japan: OR = $\exp(-1.481) = 0.23$), likely due to reduced grievances and lower perceived urgency for societal change. 
However, higher self-identified social class, whether in adolescence or adulthood, is positively associated with demonstrative action.
Individuals who self-identify as upper class both currently (\textbf{Social Class (Now): Upper}, U.S.: OR = $\exp(0.259) = 1.30$, Japan: OR = $\exp(0.703) = 2.02$) and in adolescence (\textbf{Social Class (at Age 15): Upper}, U.S.: OR = $\exp(0.260) = 1.30$, Japan: OR = $\exp(0.483) = 1.62$) are more likely to engage in demonstrative actions. 
In contrast, those from lower social backgrounds in youth (\textbf{Social Class (at Age 15): Lower}) show a reduced likelihood of participating in demonstrative actions (U.S.: OR = $\exp(-0.134) = 0.88$, Japan: OR = $\exp(-0.159) = 0.85$).
These findings suggest that while economic privilege in early life may reduce the motivation for societal change, those with consistently higher social status appears to facilitate greater willingness toward societal engagement.
One plausible explanatory mechanism is that higher and elevated social class may bolster individuals' confidence in their perspectives, potentially leading to overconfidence and assertive engagement in social movements. 
This interpretation aligns with prior research demonstrating that individuals occupying higher-paying positions display increased assertiveness and willingness to publicly express critical or oppositional viewpoints~\cite{Yamaguchi2015}.

Despite these broad similarities, notable differences emerge between the U.S. and Japan. 
\textbf{Conservative Orientation} is positively associated with demonstrative action in the U.S. (OR = $\exp(0.218) = 1.24$) but negatively associated in Japan (OR = $\exp(-0.439) = 0.64$), implying that U.S. conservatives are more inclined to protest, whereas Japan’s protesters often lean liberal. 
This patter align with broader ideological shifts: while liberals traditionally dominate protest activity across democracies~\cite{kostelka2019s}, recent evidence highlights increased mobilization among conservative groups in the U.S., especially following the COVID-19 pandemic~\cite{caren2023right}.
\textbf{Political Polarization} exerts a negative influence in the U.S. (OR = $\exp(-0.995) = 0.37$) but no significant effect in Japan, reinforcing the observations from Figure~\ref{political_poralization} that demonstrative action in the U.S. is relatively common even among politically moderate individuals.

Differences in media consumption also emerge. 
\textbf{Social Media Usage} shows no effect in the U.S. (OR = $\exp(0.012) = 1.01$) but a positive one in Japan (OR = $\exp(0.260) = 1.30$), \textbf{Message App Usage} is beneficial only in the U.S. (OR = $\exp(0.351) = 1.42$), and \textbf{Video Usage} shows a negative impact in the U.S. (OR = $\exp(-0.140) = 0.87$) but no effect in Japan (OR = $\exp(0.020) = 1.02$). 
These variations suggest that different media serve as primary mobilization channels in each country.
Notably, holding a \textbf{Postgraduate degree} in Japan is linked to a significant increase in demonstrative action  (OR = $\exp(0.673) = 1.96$), whereas Bachelor’s degree has no significant effect.

\subsubsection*{Diffusion Action in Conspiracy Theory} 
In examining how political attitudes shape actions related to conspiracy theory diffusion, common patterns in both country emerge. 
\textbf{Political polarization} strongly enhances diffusion behaviors in both the U.S. (OR = $\exp(0.689)=1.99$) and Japan (OR = $\exp(0.564)=1.76$). 
However, in contrast to its influence on demonstrative actions, \textbf{conservative orientation} shows no significant relationship with diffusion in either country (U.S.: OR = $\exp(0.018) = 1.02$, Japan: OR = $\exp(-0.024) = 0.97$), highlighting a nuanced divergence between ideological influences on demonstrative and diffusion actions.

Both countries display notable effects from socioeconomic factors. 
Individuals who had higher \textbf{Economic Capital at Age 15} (U.S.: OR = $\exp(0.241) = 1.27$, Japan: OR = $\exp(0.466) = 1.59$) and higher \textbf{Cultural Capital at Age 15} (U.S.: OR = $\exp(0.469) = 1.60$, Japan: OR = $\exp(0.685) = 1.98$) exhibit increased diffusion behaviors. 
Similarly, individuals with higher perceived social status in adolescence (\textbf{Social Class (at Age 15): Upper}, U.S.: OR = $\exp(0.045) = 1.05$, Japan: OR = $\exp(0.043) = 1.05$) and those experiencing \textbf{Upward Social Mobility} (U.S.: OR = $\exp(0.138) = 1.15$, Japan: OR = $\exp(0.302) = 1.35$) also demonstrate heightened diffusion tendencies. 
These findings suggest that early-life cultural and economic resources may equip individuals with greater confidence or skill in navigating and disseminating information, including conspiracy theories, highlighting the influence of childhood socialization on later communicative behaviors.

Media usage, whether traditional or digital platforms, facilitates diffusion behavior.
Both traditional media such as \textbf{TV/Newspaper Usage} (U.S.: OR = $\exp(0.109) = 1.12$, Japan: OR = $\exp(0.024) = 1.02$) and \textbf{Radio/Magazine Usage} (U.S.: OR = $\exp(0.090) = 1.10$, Japan: OR = $\exp(0.259) = 1.30$), as well as newer media forms including \textbf{Social Media Usage} (U.S.: OR = $\exp(0.082) = 1.09$, Japan: OR = $\exp(0.306) = 1.36$), \textbf{Video Usage} (U.S.: OR = $\exp(0.228) = 1.26$, Japan: OR = $\exp(0.114) = 1.12$), and \textbf{Online News Usage} (U.S.: OR = $\exp(0.134) = 1.14$, Japan: OR = $\exp(0.085) = 1.09$) exhibit significant effects. 
The consistently positive influence of diverse media platforms, both traditional and contemporary, highlights the generalized role of media engagement in diffusion action.

Additional demographic and attitudinal factors are linked to diffusion action in conspiracy theories; 
\textbf{Age} (U.S.: OR = $\exp(0.002) = 1.00$ per year increase, Japan: OR = $\exp(0.015) = 1.02$ per year increase), \textbf{Household Income} (U.S.: OR = $\exp(0.232) = 1.26$, Japan: OR = $\exp(0.018) = 1.02$), \textbf{Trust in Scientist} (U.S.: OR = $\exp(0.081) = 1.08$, Japan: OR = $\exp(0.053) = 1.05$), and \textbf{Having Books (Now)} (U.S.: OR = $\exp(0.059) = 1.06$, Japan: OR = $\exp(0.042) = 1.04$ exhibits a positive association, 
whereas \textbf{Male} (U.S.: OR = $\exp(-0.102) = 0.90$, Japan: OR = $\exp(-0.093) = 0.91$), \textbf{Permanent Employee} (U.S.: OR = $\exp(-0.118) = 0.89$, Japan: OR = $\exp(-0.080) = 0.92$), and \textbf{Trust in Government} (U.S.: OR = $\exp(-0.361) = 0.70$, Japan: OR = $\exp(-0.121) = 0.89$) exhibits a negative association.

As cross-national differences, \textbf{Social Class (Now): Upper} negatively influences diffusion action in the U.S. (OR = $\exp(-0.194) = 0.82$), contrasting with a positive effect in Japan (OR = $\exp(0.413) = 1.51$). 
This suggests that higher social individuals in the U.S. may suppress such dissemination actions, reflecting greater concern over reputational risks or societal norms.
Another critical difference pertains to \textbf{Religiosity}, which positively correlates with diffusion in Japan (OR = $\exp(0.657) = 1.93$), but has negligible effects in the U.S. (OR = $\exp(-0.072) = 0.93$).

\subsubsection*{Overall Patterns across the Stages of Conspiracy Engagement} 
Our analysis across the four sequential stages, Recognition, Belief, Demonstrative Action, and Diffusion Action, reveals consistent patterns as well as notable cross-national differences in conspiracy theory engagement between the U.S. and Japan. 
Several demographic and attitudinal variables exhibit consistent effects across these stages. 
Male respondents consistently demonstrate greater likelihood of recognition, belief, and demonstrative actions, although they exhibit relatively lower rates of diffusion actions. 
Political polarization consistently exhibits a positive association across all stages in Japan, whereas in the U.S., its positive influence show only in the recognition and diffusion stages. 
Conservative orientation is another consistent predictor. 
It significantly facilitates recognition and belief across both nations, reinforcing prior evidence that conspiracy theories often resonate with conservative worldviews. 
Notably, however, conservative orientation does not significantly relate to diffusion action, suggesting that political orientation facilitates exposure and acceptance but not necessarily dissemination.
Trust in government exhibits consistently negative effects on recognition, belief, and diffusion actions in both countries; however, it is uniquely positively correlated with demonstrative actions.
Religiosity demonstrates a particularly robust positive relationship with recognition, belief, and demonstrative actions across both nations, highlighting a structural affinity between religious and conspiratorial worldviews, potentially due to similar explanatory frameworks regarding uncertainty and external threats. 
However, this effect is notably weaker or absent in the context of diffusion action, suggesting that while religiosity facilitates belief internalization and public expression, it does not necessarily prompt diffusion action.
Engagement with traditional media, particularly TV/Newspaper usage, consistently acts as a protective factor, reducing conspiracy recognition, belief, and demonstrative action. 
This reinforces previous findings that traditional media consumption may limit exposure to or endorsement of conspiratorial narratives~\cite{de2021beliefs}.

Our analysis also reveals important cross-national variations in the determinants of conspiracy engagement.
Age presents divergent patterns; in the U.S., younger individuals demonstrate higher susceptibility to conspiracy belief and engagement in demonstrative actions, whereas in Japan, older individuals are more involved at these stages. 
Media usage patterns also show cultural variations. 
While radio and magazine consumption positively correlates with conspiracy belief in the U.S., it shows a negative correlation in Japan. 
Conversely, video usage strongly predicts belief in Japan but not significantly in the U.S. 
These findings highlight the need for appropriate research into how media environments influence the spread and acceptance of conspiracy theories depending on media ecosystem in each country

Our findings extend and refine existing literature, particularly regarding the roles of political polarization and online media consumption.
Previous research identified strong associations between extreme political polarization, online media usage, and conspiracy beliefs. 
However, our sequential modeling approach, accounting explicitly for recognition as an intermediate stage, reveals diminished or even reversed effects. 
It means that politically polarized individuals frequently encounter conspiracy narratives; however, this polarization may not necessarily translate into the adoption of conspiracy beliefs. 
Such results suggest that previous research overlooked earlier engagement stages and potentially overstated the direct impact of polarization on belief formation.

Lastly, socioeconomic and cultural factors significantly shape engagement trajectories across all stages. 
Early-life economic and cultural capital significantly influence individuals' likelihood of engaging with conspiracy theories; higher levels of economic and cultural capital during childhood tend to increase participation in diffusion actions, whereas lower economic capital tends to be correlate with conspiracy beliefs.
Furthermore, individuals currently identifying as upper social class are particularly inclined toward demonstrative actions, while those experiencing upward social mobility exhibit a greater tendency toward diffusion activities. 
These findings highlight the critical influence of early-life experiences and socioeconomic trajectories in shaping individual interactions with conspiracy narratives.

\section*{Discussion}\label{discussion}

\subsection*{A Multi-Stage Framework for Understanding Conspiracy Engagement}
Our study conceptualizes conspiracy theory engagement as a multi-stage process consisting of recognition, belief, demonstrative action, and diffusion action. 
By explicitly modeling each stage, we reveal intricate dynamics often overlooked in studies that treat conspiracy beliefs as singular outcomes, as evidenced by the patterns observed in our Political Polarization Score.
In particular, we demonstrate that recognition acts as a critical gatekeeper for subsequent stages, significantly determining whether individuals adopt conspiratorial beliefs or behaviors. 
This highlights the importance of distinguishing between mere exposure and deeper levels of cognitive and behavioral engagement, a distinction essential not only in conspiracy theory research but also in broader studies examining information diffusion and belief formation in digital societies.

\subsection*{Differentiating Demonstrative and Diffusion Actions}
Our findings reveal essential distinctions between two forms of conspiracy-related behaviors: demonstrative actions and diffusion actions.
Although both constitute active engagements following belief in conspiracy theories, the determinants influencing these behaviors differ significantly. 
Demonstrative actions, involving public protests or direct political contacts, often carry strong political symbolism, attracting individuals aligned with specific ideological positions. 
In contrast, diffusion actions rely predominantly on digital networks and require minimal physical or social commitment, facilitating participation from individuals broadly engaged with diverse media channels, regardless of explicit political intentions. 
We find that the determinants of these two forms of conspiracy-related actions differ in their associations with political polarization, conservative orientation, age, trust in government, media consumption patterns, and socioeconomic background. 

Specifically, demonstrative actions tend to carry more explicit political messaging, mobilizing individuals who strongly align with certain political ideologies and actively participate in offline networks~\cite{jost2018social}. 
Conversely, diffusion actions, conducted predominantly via digital platforms, may involve individuals who are less politically committed but are highly engaged with diverse media ecosystems~\cite{piat2019slacktivism,cabrera2017activism}.
Indeed, our findings reveal that while political polarization consistently drives diffusion actions across both countries, demonstrative actions are more closely associated with specific political ideologies.
Likewise, demonstrative actions are more prevalent among younger individuals, likely reflecting higher physical and social mobility, whereas diffusion actions attract a wider demographic spectrum due to the accessibility and minimal costs of digital media dissemination.

Thus, our findings highlight the importance of distinguishing between different types of conspiracy-related actions rather than treating them as a single category. 
Even actions that appear similar may be rooted in fundamentally different dynamics. 
Incorporating these differences into the modeling process allows for a more precise understanding of how individuals’ political, socioeconomic, and media-related attributes shape their distinct pathways from conspiracy beliefs to behaviors, ultimately leading to more effective analysis and intervention strategies.

\subsection*{The Influence of Socioeconomic and Cultural Context} 
Our results highlight the critical role of early-life socioeconomic and cultural conditions as key determinants shaping conspiracy engagement.
Early-life cultural and economic resources appear to strongly influence how individuals interact with conspiracy narratives, suggesting that childhood environments play a formative role beyond mere cognitive or attitudinal predispositions. 
Specifically, individuals who experienced economic stability or had greater cultural capital in childhood exhibit a heightened propensity toward diffusion actions.
Conversely, current perceptions of lower socioeconomic status appear to make individuals more susceptible to conspiracy beliefs, potentially reflecting their economic dissatisfaction or experiences of marginalization, consistent with previous research~\cite{casara2022impact, jetten2022economic}.
Interestingly, we observe that individuals identifying as upper social class tend to engage significantly in demonstrative actions, suggesting that contemporary socioeconomic privilege can facilitate more publicly visible and organized forms of conspiratorial behavior. 
This indicates that social class not only shapes cognitive engagement with conspiracy theories but also influences the nature and visibility of subsequent actions.
These findings suggest that conspiracy engagement trajectories are deeply influenced by socioeconomic contexts experienced both in childhood and adulthood, emphasizing the necessity of incorporating life-course perspectives in conspiracy research.

\subsection*{Limitations and Implications for Future Research}

Several methodological considerations and limitations of this study should be addressed when interpreting our findings. 
First, our analyses are based on self-reported survey data collected online, potentially subjecting results to biases such as social desirability, recall inaccuracies, and respondents' subjective interpretations of conspiracy-related behaviors. 
Although we applied rigorous quality controls to mitigate these concerns, self-report measures inherently risk inflating or underestimating actual engagement, especially regarding socially sensitive behaviors like public demonstrations or conspiracy dissemination.
Second, our cross-sectional research design limits causal inference, leaving open the question of temporal or causal directionality. 
It remains unclear whether factors such as political attitudes, trust levels, or media habits drive conspiracy engagement or whether engagement itself reinforces or reshapes these characteristics. 
Future studies would benefit from employing longitudinal or experimental designs capable of clarifying these causal pathways and their temporal order.

Additionally, the selected conspiracy theories examined in this study represent only a subset of the broader conspiracy landscape, which encompasses diverse themes and contexts. 
The generalizability of our findings to other conspiracy narratives or cultural settings remains uncertain.
As our results reveal pronounced cross-national variations—such as differing age-related patterns and media effects between Japan and the U.S.—the specific cultural, historical, and institutional contexts underlying these differences require further investigation. 
Indeed, despite both being economically developed democratic nations, the U.S. and Japan exhibit distinct engagement patterns, emphasizing the complexities involved in international conspiracy theory interventions.
Given these observed cross-national variations, there is a pressing need for research that extends beyond single-country analyses, such as studies examining conspiracy theory dynamics across 26 countries~\cite{imhoff2022conspiracy}. 
Future comparative studies encompassing a wider range of cultural and structural contexts would provide deeper insights into the global mechanisms shaping conspiracy beliefs and behaviors. 
Such broader investigations are essential for developing culturally adaptive and effective strategies to mitigate the spread of conspiracy theories and their societal consequences.





\section*{Conclusion}\label{conclusion}

Our study provides a comprehensive analysis of conspiracy theory engagement by conceptualizing it as a multi-stage process encompassing recognition, belief, demonstrative action, and diffusion action. 
By applying a Bayesian hierarchical model, we demonstrate that conspiracy engagement is not a singular phenomenon but a sequential progression shaped by distinct social, political, and economic determinants at each stage. 
This approach accounts for the role of recognition as a crucial gateway, filtering individuals who later adopt beliefs and engage in conspiratorial behaviors, thereby offering a more precise and structured framework for modeling the dynamics of conspiracy theory engagement.

Our findings also reveal fundamental distinctions between demonstrative and diffusion actions. 
While demonstrative actions require greater commitment and are more common among younger individuals, those with higher social status, and those aligned with specific political ideologies, diffusion actions engage a broader demographic, including older individuals and those with greater economic and cultural capital who actively consume diverse media sources.
These differences show the necessity of distinguishing between forms of conspiracy-related behaviors rather than treating them as a singular phenomenon.

These insights hold practical implications for strategies aimed at mitigating the spread of conspiracy theories. 
Recognizing the multi-stage and multifaceted nature of conspiracy engagement highlights the necessity of targeted interventions responsive to specific stages and behaviors
Early-stage interventions might concentrate on strengthening media literacy and critical thinking to reduce the recognition and belief. 
Addressing later-stage behaviors, such as demonstrative and diffusion actions, requires differentiated strategies. For instance, interventions targeting broader, digitally engaged populations can help restrain the large-scale spread of conspiracy content.
By adopting a multi-stage framework, stakeholders including policymakers, educators, and media platforms can more effectively design, implement, and refine interventions that address the distinct drivers at each stage. 
This approach enables the development of more targeted and context-sensitive strategies to mitigate the societal impact of conspiracy theories.

\section*{Methods}\label{method}

\subsection*{Survey design and participants}
To investigate how conspiracy theories are recognized, believed, and acted upon in Japan and the U.S., an online survey was conducted by Cross Marketing Inc. between September 25 and October 4, 2024, targeting adult participants (aged 20–69) in both countries.
Questionnaires were randomly distributed to monitors registered with the company and its partner organizations. 
A total of 20,150 responses were collected in Japan and 19,783 in the U.S., with quotas set by gender and age to approximate national census distributions. 

Quality control measures included screening questions and checks for response consistency. 
Respondents who failed trap questions or completed the survey in unrealistically short times were excluded, resulting in final sample sizes of 16,693 in Japan and 13,578 in the U.S. (Tables S1 and S2 in the Supplementary Information).

\subsection*{Data collection}
We conducted an online survey to investigate how individuals in Japan and the U.S. recognize, believe, and act upon conspiracy theories. 
The questionnaire collected demographic characteristics (age, gender, educational attainment, household income, and employment status), political orientation, trust in government and scientists, religiosity, media usage habits, and social capital. 
Respondents were also asked about their engagement with 11 pre-identified conspiracy theories in their respective countries across three sequential stages: recognition, belief, and actions.

To capture the sequential stages of engagement with conspiracy theories, respondents were first asked whether they had ``heard of or encountered'' each conspiracy theory  (\textbf{recognition}). 
For each theory they recognized, respondents indicated their \textbf{belief} (``I believe it is true,'' ``I don’t know,'' or ``I believe it is false''.) 
Subsequently, respondents who reported believing at least one conspiracy theory were asked whether they had engaged in \textbf{demonstrative actions} (e.g., attending demonstrations, contacting relevant parties) or \textbf{diffusion actions} (e.g., sharing or disseminating the conspiracy information). 
Because a later stage (e.g., demonstrative action) could not be true unless the preceding stage (e.g., belief) was true, the effective sample size for each outcome depends on the participant’s responses at earlier stages.

Table~\ref{variable_table} provides an overview of these explanatory variables used in our analyses, along with details on variable transformations for both Japan and the U.S.
The survey questions and the construction of variable values are described in Appendix A and B of the Supplementary Information, while distributional details appear in Figures S3 - S8.

\begin{table}[ht]
\centering
\caption{Overview of Explanatory Variables}
\label{variable_table}
\begin{tabular}{p{4cm} p{9cm}}
    \toprule
     \multicolumn{1}{c}{\textbf{Variable}} &  \multicolumn{1}{c}{\textbf{Description}} \\
    \midrule
    \multicolumn{2}{c}{\textbf{Demographics}} \\
    \cmidrule{1-2}
    Age & Years beyond age 20 (both countries). \\
    Male & 1 if male; 0 otherwise. \\
    Marital Status & 1 if married; 0 otherwise. \\
    Urbanization & \parbox{9cm}{In the US: 1 = Resident of a large city (over 250,000); 0 = Otherwise. \\In Japan: 1 = Resident of a government-designated city; 0 = Otherwise.} \\    
    Bachelor's degree & 1 if holding a bachelor's degree or higher; 0 otherwise. \\
    Postgraduate Degree & 1 if holding a postgraduate degree; 0 otherwise. \\
    \midrule

    \multicolumn{2}{c}{\textbf{Social Status}} \\
    \cmidrule{1-2}
    Household Income & Normalized so that 10,000,000 JPY (Japan) or \$100,000 (US) equals 1. \\
    Student & 1 if a student; 0 otherwise. \\
    Permanent Employee & 1 if employed as a permanent employee; 0 otherwise. \\
    Large Company & 1 if employed at a firm with at least 1000 employees; 0 otherwise. \\
    Medium-sized Company & 1 if employed at a firm with 100–999 employees; 0 otherwise. \\
    Small Company & 1 if employed at a firm with 10–99 employees; 0 otherwise. \\
    \midrule
    
    \multicolumn{2}{c}{\textbf{Political Orientation}} \\
    \cmidrule{1-2}
    Political Polarization & Continuous score (0–1) indicating extremity (0 = moderate, 1 = extreme). \\
    Conservative Orientation & Continuous score where 1 indicates conservative and 0 indicates liberal. \\
    \midrule
    
    \multicolumn{2}{c}{\textbf{Trust}} \\
    \cmidrule{1-2}
    Trust in Government & Normalized score (0–1) for government trust (1 = high). \\
    Trust in Scientist & Normalized score (0–1) for trust in scientists (1 = high). \\
    Religiosity & Normalized score (0–1) for religious commitment (1 = high). \\
    \midrule

    \multicolumn{2}{c}{\textbf{Media Usage Habits}} \\
    \cmidrule{1-2}
    Social Media Usage & 1 if daily usage is at least 30 minutes (e.g., X, Instagram, Facebook), 0 otherwise. \\
    Video Usage & 1 if daily usage is at least 30 minutes (e.g., YouTube, Netflix), 0 otherwise. \\
    TV/Newspaper Usage & 1 if daily usage is at least 30 minutes, 0 otherwise. \\
    Radio/Magazines Usage & 1 if daily usage is at least 30 minutes, 0 otherwise. \\
    Online News Usage & 1 if daily usage is at least 30 minutes, 0 otherwise. \\
    Message App Usage & 1 if daily usage is at least 30 minutes (e.g., Messenger, Line), 0 otherwise. \\
    Personal Website Usage & 1 if daily usage is at least 30 minutes, 0 otherwise. \\
    \midrule
    
    \multicolumn{2}{c}{\textbf{Cultural, Economic and Social Capital}} \\
    \cmidrule{1-2}
    Cultural Capital (at Age 15) & Number of cultural items available at age 15, normalized to a maximum of 1. \\
    Economic Capital (at Age 15) & Number of economic items available at age 15, normalized to a maximum of 1. \\
    Having Books (Now) & Number of books currently available, normalized so that 100 books equal 1. \\
    Having Books (at Age 15) & Number of books at age 15, normalized so that 100 books equal 1. \\
    Reading Books & Number of books read per year, normalized so that 10 books equal 1. \\
    Number of Friends & Number of friends, normalized so that 100 friends equal 1. \\ 
    Bachelor's degree (Parents) & 1 if at least one parent holds a bachelor's degree or higher; 0 otherwise. \\
    Postgraduate Degree (Parents) & 1 if at least one parent holds a postgraduate degree or higher; 0 otherwise.\\
    
    Social Class (Now): Upper & 1 if self-perceived as upper class, 0 otherwise. \\
    Social Class (Now): Lower & 1 if self-perceived as lower class, 0 otherwise. \\
    Social Class (at Age 15): Upper & 1 if perceived as upper class at age 15, 0 otherwise. \\
    Social Class (at Age 15): Lower & 1 if perceived as lower class at age 15, 0 otherwise. \\
    Upward Social Mobility & 1 if current social class is perceived as higher than at age 15, 0 otherwise. \\
    \bottomrule
\end{tabular}
\end{table}

\subsection*{Analysis Method}
The purpose of our analysis is to construct a regression model that explains how individuals come to recognize conspiracy theories, develop beliefs in them, and subsequently engage in demonstrative and diffusion actions. 
By modeling these processes as interrelated stages, we aim to capture both direct and indirect pathways through which explanatory factors influence recognition, belief formation, and two forms of behavioral responses.

\noindent \textbf{Conceptual Structure}

\begin{figure}[t]
    \centering
    \includegraphics[width=\linewidth]{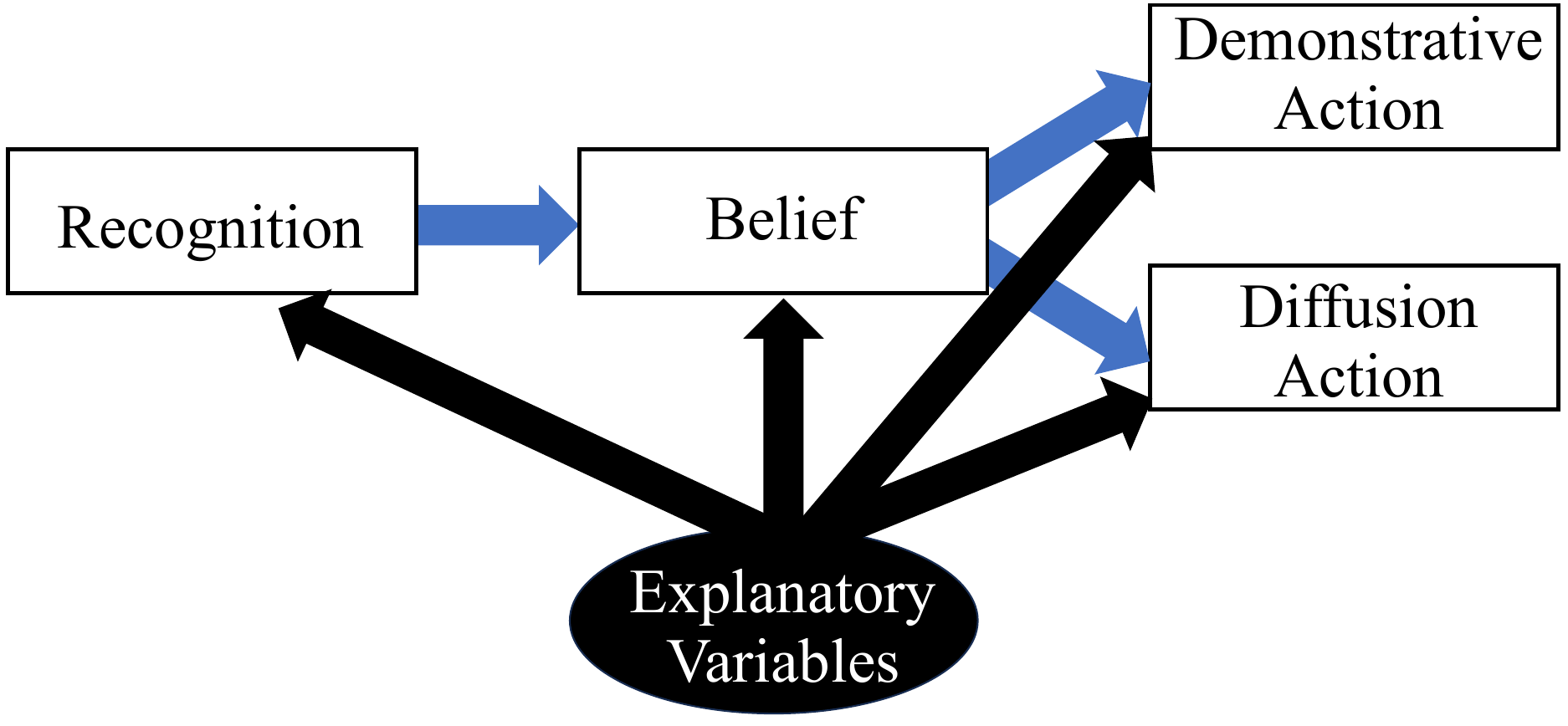}
    \caption{
    This figure depicts the hypothesized pathways by which individual attributes influence the process of conspiracy theory engagement. 
    The process begins with Recognition (awareness of the conspiracy theory), followed by Belief (accepting the theory as true if recognized), and finally leads to two forms of action: Demonstrative Action and Diffusion Action. 
    Each stage is influenced both directly by explanatory variables and indirectly through mediation by prior stage.
    }
    \label{Concept_image}
\end{figure}

Figure~\ref{Concept_image} illustrates the conceptual path diagram.
Recognition of conspiracy theories is influenced by a set of explanatory variables.
Belief in conspiracy theories depends both on recognition and the same explanatory variables. 
Demonstrative action, such as participating in visible public demonstrations, is posited to be influenced by belief and the explanatory variables.
Diffusion action, referring to efforts to spread or disseminate conspiracy-related information on social media, is also influenced by belief and the explanatory variables.
The diagram indicates how recognition, belief, and two forms of action (demonstrative and diffusion) are modeled in sequence, reflecting both direct and indirect pathways of influence from explanatory variables.

\noindent \textbf{Hierarchical Bernoulli Model}

We model all four outcomes, recognition ($R$), belief ($B$), demonstrative action ($M$), and diffusion action ($F$), as Bernoulli-distributed responses. 
Each outcome is specified with a logistic link function to estimate the probability of occurrence at each stage. 
By jointly modeling these outcomes, we account for correlations in the error structure and better capture mediation, like pathways among the outcomes:

\begin{align*}
R &\sim \text{Bernoulli}\left( \sigma(\alpha_r + \bm{\beta_r X}) \right), \\
B &\sim \text{Bernoulli}\left( \sigma(\alpha_b + \beta_{b1} R + \bm{\beta_{b2} X}) \right), \\
M &\sim \text{Bernoulli}\left( \sigma(\alpha_m + \beta_{m1} B + \bm{\beta_{m2} X}) \right), \\
F &\sim \text{Bernoulli}\left( \sigma(\alpha_f + \beta_{f1} B + \bm{\beta_{f2} X}) \right).
\end{align*}

\noindent where $R$ represents the probability of recognition of a conspiracy theory, $B$ represents the probability of belief in a conspiracy theory conditioned on $R$, $M$ represents demonstrative action influenced by $B$, and $F$ represents diffusion action, which is also dependent on $B$. 
The vector $\bm{X}$ includes explanatory variables shared across all models, while $\alpha$ and $\bm{\beta}$ are coefficients to be estimated. 
The logistic (sigmoid) function $\sigma(\cdot)$ ensures probabilistic modeling.

By jointly estimating these equations within a Bayesian framework, we capture both direct effects of explanatory variables ($\bm{X}$) on each stage and indirect effects mediated by prior stages. 
The direct effects of $\bm{X}$ correspond to their immediate influence on recognition, belief, demonstrative action, and diffusion action. 
The indirect effects arise as recognition influences belief, which in turn conditions the likelihood of demonstrative and diffusion actions. 
Since recognizing a conspiracy theory is a prerequisite for belief, individuals' recognition levels significantly shape their likelihood of developing a belief. Similarly, belief serves as a necessary condition for engaging in either demonstrative or diffusion actions. 
By analyzing the relationships between estimated coefficients, we can quantify both direct and mediated pathways through which explanatory factors exert influence. 
This allows us to determine not only the individual effects of recognition and belief but also how these stages function as crucial gateways to subsequent behaviors.

\noindent \textbf{Implementation Details}

For Bayesian inference, we employed four Markov chains, each with 2,000 iterations, discarding the first 1,000 as warm-up, using R (v4.2.2) with the \texttt{brms} package (v2.20.4).
To mitigate divergent transitions, we set $\text{adapt\_delta} = 0.95$.
Convergence was confirmed through trace plots, effective sample sizes, and Gelman–Rubin diagnostics.
Full details of the posterior predictive checks, the convergence diagnostics and leave-one-out cross-validation results are provided in Appendix C in the Supplementary Information.

\subsection*{Alternative Model Specifications} 
To verify the robustness of our findings, we also explored alternative modeling approaches, including Structural Equation Modeling (SEM) and Sequential Generalized Linear Models (Sequential GLM). 
While the SEM estimates appeared unstable, particularly at the later stages of action, the results from the Sequential GLM were largely consistent with those obtained from our primary Bayesian hierarchical analysis. 
A detailed comparison of these methods and their respective estimates is provided in Appendix D in the Supplementary Information, highlighting the consistency of key findings across different analytical frameworks.

\subsection*{Data availability}
De-identified survey data and replication codes will be made available on ****.
The Supplementary Information is available at \url{https://github.com/hkefka385/threestage_conspiracy_interview/blob/main/NHB2025_ConspiracyInterview_supplementary.pdf}.

\section*{Acknowledgment}
This work was supported by Research Institute of Science and Technology for Society, Japan, Grant Number JPMJRS23L4.
We would like to express our gratitude to the following two individuals for their cooperation in this study: Atsushi Ueshima (Keio University) and Shiori Hironaka (Kyoto University).


\bibliography{sn-bibliography}


\end{document}